\documentclass[prd,singlecolumn]{revtex4}
\pdfoutput=1
\usepackage{amssymb,latexsym}
\usepackage{amsmath,amsbsy,bbm}
\usepackage{ifpdf}
\usepackage{epsfig,bm}
\usepackage{graphicx,comment}
\usepackage{color}
\usepackage{soul}
\usepackage{mathtools}
\usepackage{comment}
\usepackage[normalem]{ulem}
\begin{document}

\title{The phase transitions of the frustrated $J_1$-$J_2$ Ising model on the honeycomb lattice}
\author{Shang-Wei Li}
\affiliation{Department of Physics, National Taiwan Normal University,
  88, Sec.4, Ting-Chou Rd., Taipei 116, Taiwan}
\author{Yuan-Heng Tseng}
\affiliation{Department of Physics, National Taiwan Normal University,
  88, Sec.4, Ting-Chou Rd., Taipei 116, Taiwan}
\author{Fu-Jiun Jiang}
\email[]{fjjiang@ntnu.edu.tw}
\affiliation{Department of Physics, National Taiwan Normal University,
88, Sec.4, Ting-Chou Rd., Taipei 116, Taiwan}

\begin{abstract}
  We study the phase transitions of the frustrated $J_1$-$J_2$ Ising model on the honeycomb lattice
  using the non-perturbative first principle Monte Carlo simulations.
  Here $J_1 < 0$ and $J_2 > 0$ are the nearest and next-to-nearest couplings, respectively. In particular, the values of
  $J_2/|J_1| = 0.20, 0.22, 0.23, 0.24, 0.3, 0.5, 0.8,$ and 1.0 are considered in our study. Based on the numerical
  outcomes, we find that the phase transitions for $J_2/|J_1| = 0.20, 0.22, 0.23,$ and 0.24 are
  second order and are governed by the 2D Ising universality class. In addition, we find evidence to support the facts
  that there are transitions for $J_2/|J_1| = 0.5, 0.8$ and 1.0 and these phase transitions are
  second order. Our results also indicate phase transition is unlikely to take place for $g=0.3$. We are not able to obtain results for  $J_2/|J_1|$ $\in$ (0.24, 0.3)
  because the associated integrated autocorrelation times or (and) the equilibrium times are extremely large at the low-temperature region. A comparison between the outcomes presented here
  and the available results in the literature is briefly conducted as well.

\end{abstract}

\maketitle

\section{Introduction}

Competing interactions inside matter often lead to the occurrence of exotic states which may have great potential in applied material science. Consequently, studying models with competing interactions
is an active topic in theoretical physics \cite{Die13}.
  
Frustrated $J_1$-$J_2$ Ising model on the square lattice is a representative model with competing interactions. Here $J_1 < 0$ ($J_2 > 0$) is the interaction between any pair of nearest (next-to-nearest)
neighboring Ising spins. This model has attracted a lot of attention both analytically and numerically during the last several decades \cite{Bin80,Oit81,Lan85,Oit87,Lop93,Lop931,Alv03}, and its phase diagram as a function of $g = J_2/|J_1|$ is well-known to certain extent \cite{Kal08,Jin12,Kal12,Jin13,Kel19,Hu21,Li21,Yos23,Li24}.
For instance, at zero temperature, a phase transition from the ferromagnetic phase to the striped phase takes place at $g=0.5$.
In addition, for a given fixed $g < 0.5$, phase transition belonging to the two-dimensional (2D) Ising universality class occurs when one moves from low-temperature region to high-temperature region. For $g > 0.5$.
there is a special point $g*$ (The value of $g*$ is still under debating \cite{Jin12,Hu21,Yos23,Li24}) where the phase transitions of 
$0.5 < g < g*$ is first order, and for $g \ge g*$, the associated critical phenomena are governed by the Ashkin-Teller weak universality class. In summary, this model is well-studied and its properties are
understood fairly well.

Compared to the frustrated $J_1$-$J_2$ Ising model on the square lattice, much less is known for the
characteristics of the model having the same Hamiltonian, but on the honeycomb lattice \cite{Kat86,Bob16,Zuk19,Zuk21,Ace21}. At the moment, it is established that at zero temperature, the ferromagnetic phase exists for $g < 0.25$ and no long-range order is found for $g>0.25$. Theoretically,
although the phase transitions of $g < 0.25$ are expected to be second order and belong to
the 2D Ising universality class, this has not been confirmed by
exact numerical methods such as the Monte Carlo (MC) simulations for $0.20 < g < 0.25$. Finally, there is no definite answer regarding
whether phase transitions can be observed for any value of $g > 0.25$.

Due to these mentioned facts, in this study, we explore the phase diagram of the 2D frustrated $J_1$-$J_2$ Ising model on the honeycomb lattice using the first-principles non-perturbative Monte Carlo simulations.

By conducting a careful finite-size scaling analysis, we find convincing numerical evidence that the phase transitions for $g=0.20, 0.22, 0.23$ and 0.24 are second order and belong to the 2D Ising universality class. In addition, our results also support the scenario
that phase transitions do occur for $g=1.0, 0.8$ and 0.5. Finally,
for $g=0.3$, we conclude that it is of high possibility that phase
transition does not take place.

We would like to point out that in our investigation, we find that
the integrated autocorrelation time or the equilibrium time is extremely large for the low-temperature region of $ 0.24 < g < 0.3$. As a result, we are not
capable of obtaining information regarding whether there is any phase transition for $0.24 < g < 0.3$. 

Finally, we compare our conclusions with those available in the literature, namely Refs.~\cite{Zuk19,Zuk21,Ace21}. Although we agrees
with some of the claims of Refs.~\cite{Zuk19,Zuk21}, we find the data
presented in these references may overlook the associated integrated autocorrelation time, and hence the related error bars are likely underestimated.

It should be emphasized that there is a correspondence between the cases of $J_1 < 0$ and $J_1 > 0$ \cite{Civ25}. As a result, the phase diagram as well as the energy-related quantities should be the same for models with $J_1 > 0$ and $J_1 < 0$. The nature of the ground states, of course, depends on the sign of $J_1$. Here we will use the convention $J_1 < 0$.

The rest of the paper is summarized as follows. After the brief introduction, the model and the calculated quantities are introduced in Sect. II.
Then in Sect. III, we present our numerical data. In particular the data supporting the claims outlined in the Sect. " Introduction " are
demonstrated in detail. Finally, in Sect. IV, we give our conclusions regarding the findings of this investigation. Two appendices are added as well to show the extremely large integrated autocorrelation times and 
the equilibrium times for $0.24 < g < 0.3$ as well as an explanation of
one technical usage of the MC simulations.

\section{The considered model}

\begin{figure}
    \includegraphics[width=0.3\textwidth]{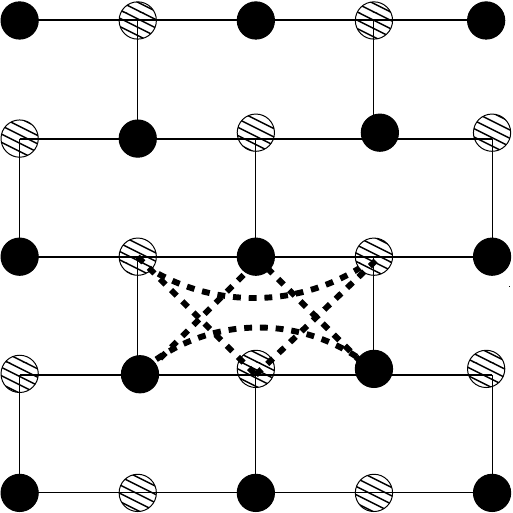}
	\caption{The frustrated $J_1$-$J_2$ Ising model on a 4 by 4 honeycomb lattice. The thin solid and thick dashed lines represent the couplings $J_1$ and $J_2$, respectively. The couplings $J_2$ are only demonstrated in a unit of the whole honeycomb lattice.}
	\label{j1j2}
\end{figure}

The Hamiltonian $H$ of the 2D frustrated $J_1$-$J_2$ Ising model on the honeycomb
lattice studied here is expressed as \cite{Bob16,Zuk19,Zuk21,Ace21}
\begin{equation}
 H = J_1 \sum_{\left< ij\right>} \sigma_i\sigma_j + J_2 \sum_{\left<\left< lm\right>\right>}\sigma_l\sigma_m,
\label{eqn}
\end{equation}
where $\sigma_i = \pm 1$ is the Ising spin at site $i$. 
Fig.~\ref{j1j2} is the pictorial representation of the model studied here. 
In the figure, the solid and dashed lines stand for the couplings $J_1 < 0$ and
$J_2 > 0$, respectively, and the coupling $J_2$ is only demonstrated in a unit of the whole honeycomb lattice.
In our calculations, we have set $J_1 = -1$ and have considered the phase
transitions for several fixed $J_2 > 0$. For convenience, we define $g$
to be $J_2/|J_1|$.

The quantities determined in this study are the susceptibility $\chi$, the specific
heat $C_v$ and the energy density $E$ (energy per site).

The susceptibility $\chi$ is defined as
\begin{equation}
\chi = \frac{N}{T}\left( \langle m^2\rangle - \langle m \rangle^2 \right),
 \end{equation} 
where $m$ is expressed by
\begin{equation}
m = \frac{1}{N} \Big|\sum_{i=1}^{N} \sigma_i\Big|.
\end{equation}
Here $N = L^2$ with $L$ being the linear system size.

The specific heat $C_v$ is given by
\begin{equation}
\frac{N}{T^2}\left( \langle E^2\rangle -\langle E\rangle^2 \right),
\end{equation}  
where $E$ and $T$ are the energy density and temperature, respectively.

\section{Numerical Results}

To investigate the proposed phase transitions of the 2D frustrated $J_1$-$J_2$
Ising model on the honeycomb lattice, we have carried out large scale Monte
Carlo calculations (MC) using the single spin flip Metropolis algorithm \cite{Met53,Has70,New99}.
First of all, we will focus on the cases of $g < 0.25$.
In particular, convincing evidence will be presented to show that
the phase transitions for $g = 0.20, 0.22, 0.23, 0.24$ are
second order and belong to the 2D Ising universality class. 

Second, the outcomes associated with $g > 0.25$ will be demonstrated as well. We will show that it is likely there are second order phase transitions for $g = 1.0, 0.8, 0.5$. For $g = 0.3$, our data implies
there is no phase transition, although data at low-temperature region with large $L$
may be needed to confirm this.

For the values of $g$ which lie between 0.24 and 0.3, we are not able to obtain any clues regarding the related phase transitions due to
the extremely long integrated autocorrelation times and equilibrium times at the low-temperature region for these $g$.

\subsection{The phase transitions for $g < 0.25$}

\subsubsection{Specific heat $C_v$}

To determine the nature of the considered phase transitions, particularly to calculate the corresponding critical temperature for a given $g$, we compute the specific heat $C_v$ as a function of $T$. For each considered $g$ and for a fixed $L$, the temperature where $C_v$ takes the maximum value can be considered as the associated pseudo-critical temperature $T_c(g,L)$.

\begin{figure}
	\includegraphics[width=0.5\textwidth]{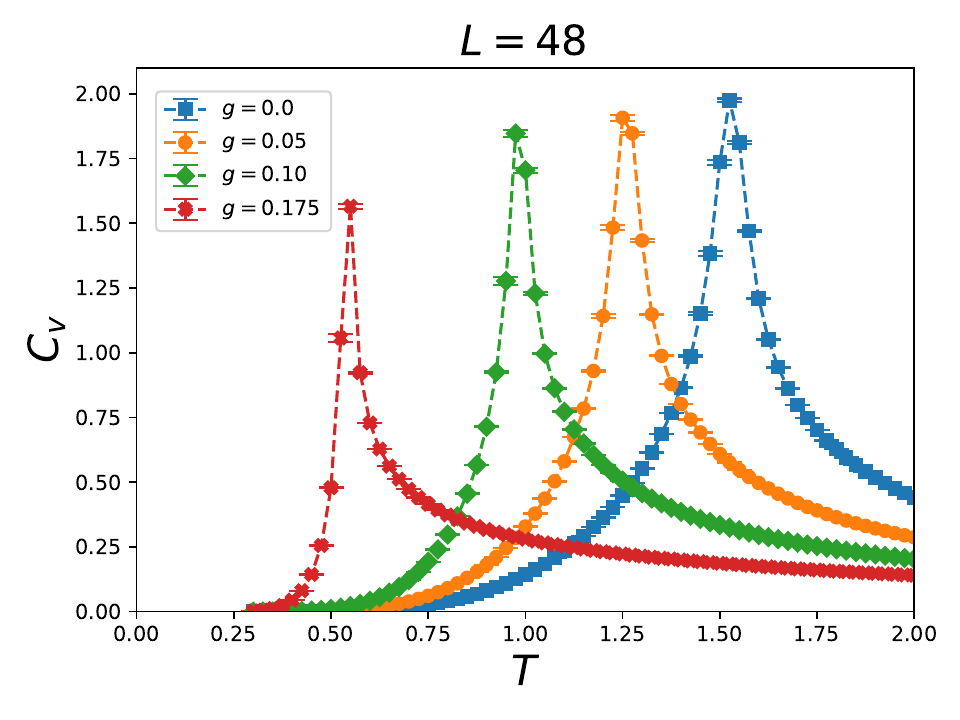}
	\caption{$C_v$ as functions of $T$ for several values of $g <0.25$ and $L=48$.}
	\label{cv1}
\end{figure}

Figure~\ref{cv1} shows the $C_v$ as functions of $T$ for $g = 0.0, 0.05, 0.1$, and 0.175 with $L=48$. The ferromagnet-like configuration is employed as the initial configuration to start the MC simulations. For $g=0.0, 0.05, 0.1$ ($g = 0.175$),
the used thermalization is $10^6$ ($2 \times 10^6$) Monte Carlo sweeps (One Monte Carlo sweep is defined as performing the spin flip algorithm for all the spins once). In addition, we have 1000 (10000) data points for $g=0.0, 0.05, 0.1$ ($g=0.175$) and each data point is the average of results from conducting 5000 MC sweeps. The errors are estimated by
multiplying the naive errors by the squared root of the integrated autocorrelation time $\tau_{\text{int}}$. We use the package emcee to obtain the $\tau_{\text{int}}$ of a MC series \cite{For13}. When applying the emcee to analyze a MC series, it will return the associated $\tau_{\text{int}}$ without any warning message only in the case that the length of the series divided by
50 is greater $\tau_{\text{int}}$.

As can be seen from the figure,
for each $g$, a maximum value of $C_v$ does appear. The results demonstrated in fig.~\ref{cv1} are in good agreement with that of
Ref.~\cite{Zuk21}. This provides convincing numerical outcomes to support the correctness of
our MC codes.

\begin{figure}
	\begin{center}
		\hbox{
			\includegraphics[width=0.45\textwidth]{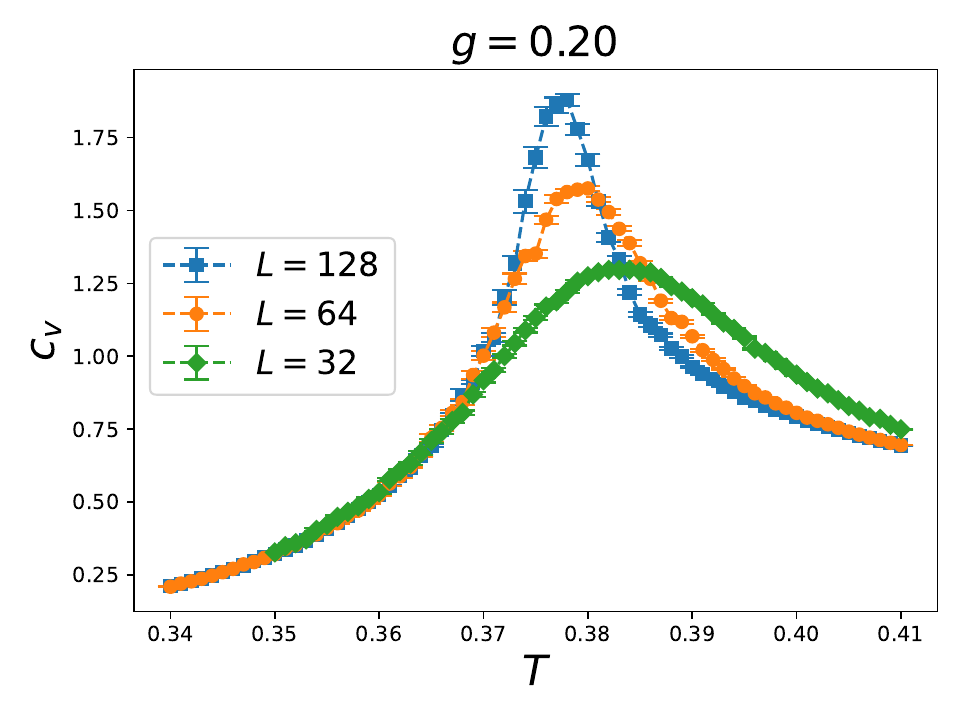}
			\includegraphics[width=0.45\textwidth]{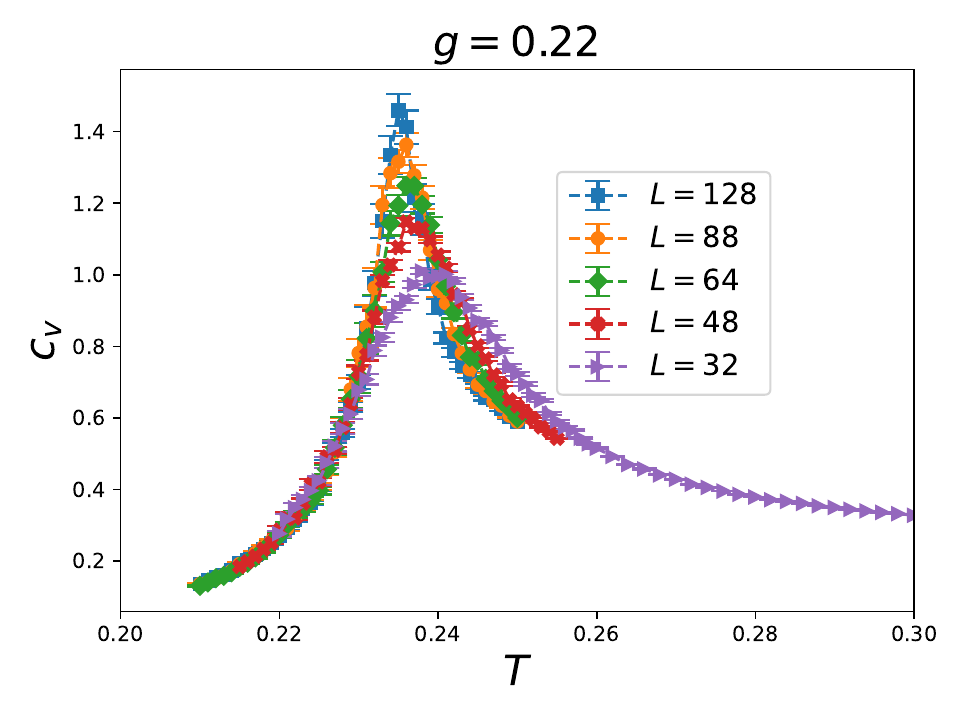}
		}
	\end{center}
	\caption{$C_v$ as functions of $T$ for several $L$ for $g=0.20$ (left) and $g = 0.22$ (right).}
	\label{cv4}
\end{figure}

\begin{figure}
	\begin{center}
		\hbox{
			\includegraphics[width=0.45\textwidth]{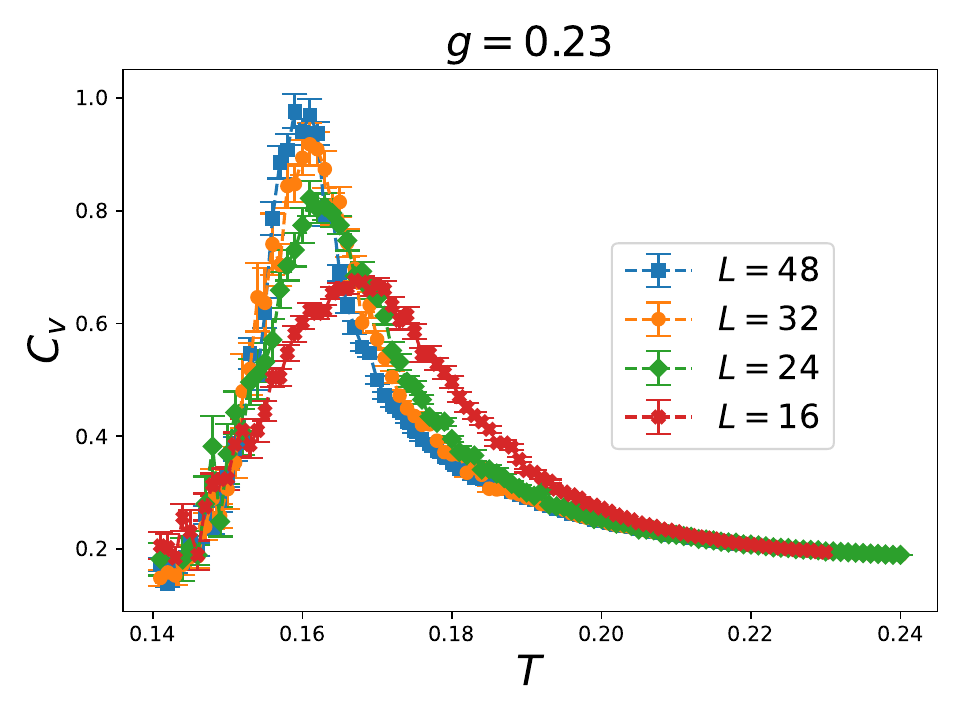}
			\includegraphics[width=0.45\textwidth]{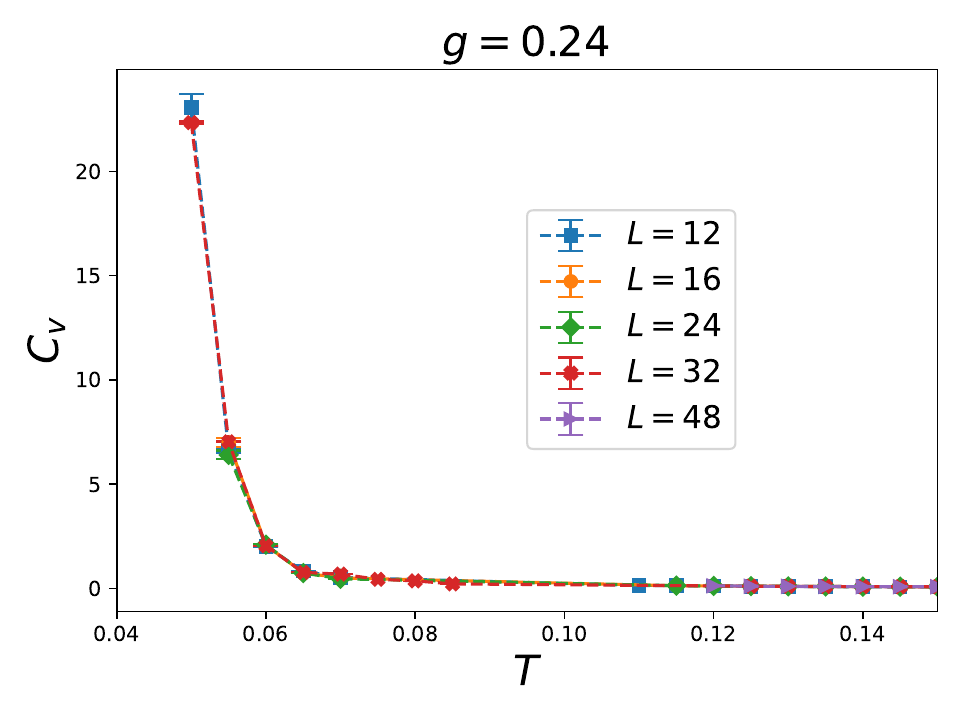}
		}
	\end{center}
	\caption{$C_v$ as functions of $T$ for several $L$ for $g=0.23$ (left) and $g = 0.24$ (right).}
	\label{cv5}
\end{figure}

After verifying the correctness of our codes, we turn to studying
the phase transitions close to $g=0.25$.
$C_v$ as functions of $T$ for various $L$ for $g=0.20,0.22,0.23$, and 0.24 are shown as the left and the right panels of figs.~\ref{cv4} 
and \ref{cv5}, respectively. Every data shown in these figs. is obtained from few thousand to a few ten thousand data points (and each of these data points is the averaged results of performing several thousand to few ten thousand MC sweeps). In addition, even in the case that the emcee returns a result without a warning message, only those data with their $\tau_{\text{int}}$ smaller than some reasonable numbers are accepted (and presented).
For different physical quantities and (or) $g$, the used numbers may be varied since the associated correlations could be different.

The outcomes demonstrated in the figures indicate the heights of the peaks of $C_v$ have the trend of saturating to a constant with increasing $L$ for $g=0.20,0.22$ and 0.23. Since the maximum of $C_v$
at finite lattices $L$ scale as $L^{\alpha/\nu}$, the observed phenomenon implies $\alpha = 0$ which is consistent with the 2D Ising universality class. 

For $g=0.24$, we are not able to obtain convergent results with $T \le 0.05$. Therefore, conclusive claim cannot be drawn from the quantity $C_v$. Later, when the observables $\chi$ and $m$ are concerned,
evidence supporting the phase transition is second order and belongs to the 2D Ising universality class will be given.

We would like to emphasize the fact that the $T_c$ of $g=0.23$ obtained in Ref.~\cite{Zuk21} agrees well with ours. Indeed, the peak of fig.~\ref{cv5} for $g=0.23$ indicates $T_c \sim 0.16$. In Ref.~\cite{Zuk21}, $E$ of $g=0.23$ is determined by
the method of temporal annealing. Their result show a sudden rise in $E$ around $T \sim 0.15$, implying the presence of a phase transition close to $T \sim 0.15$. This is consistent with our result of $T_c \sim 0.16$ for $g=0.23$.

\subsubsection{Magnetization $m$}

\begin{figure}
\begin{center}
	\hbox{
		\includegraphics[width=0.45\textwidth]{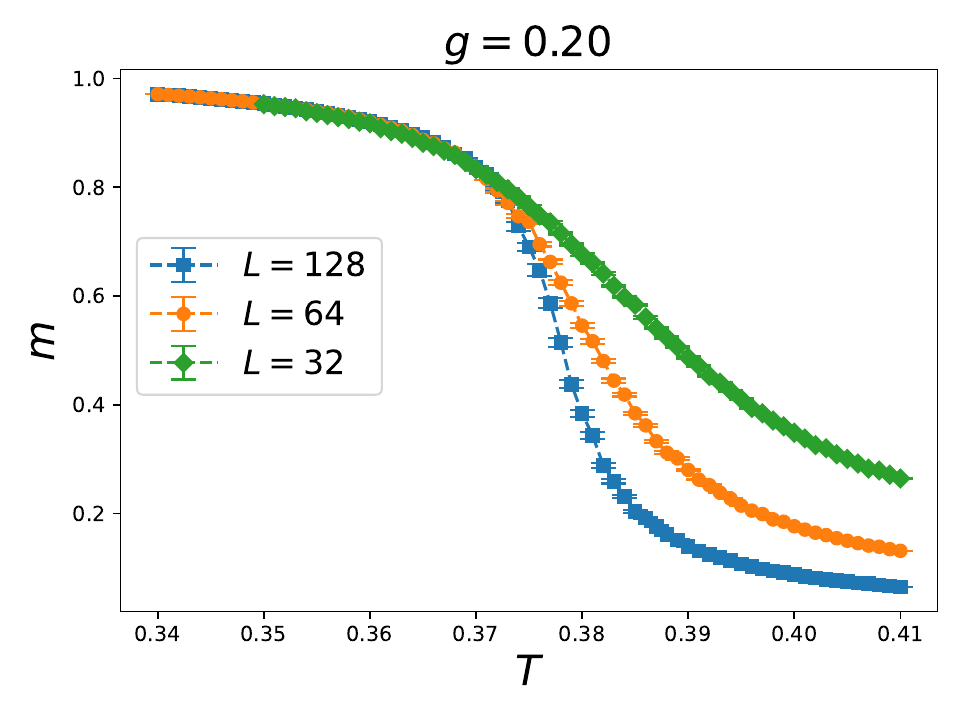}
		\includegraphics[width=0.45\textwidth]{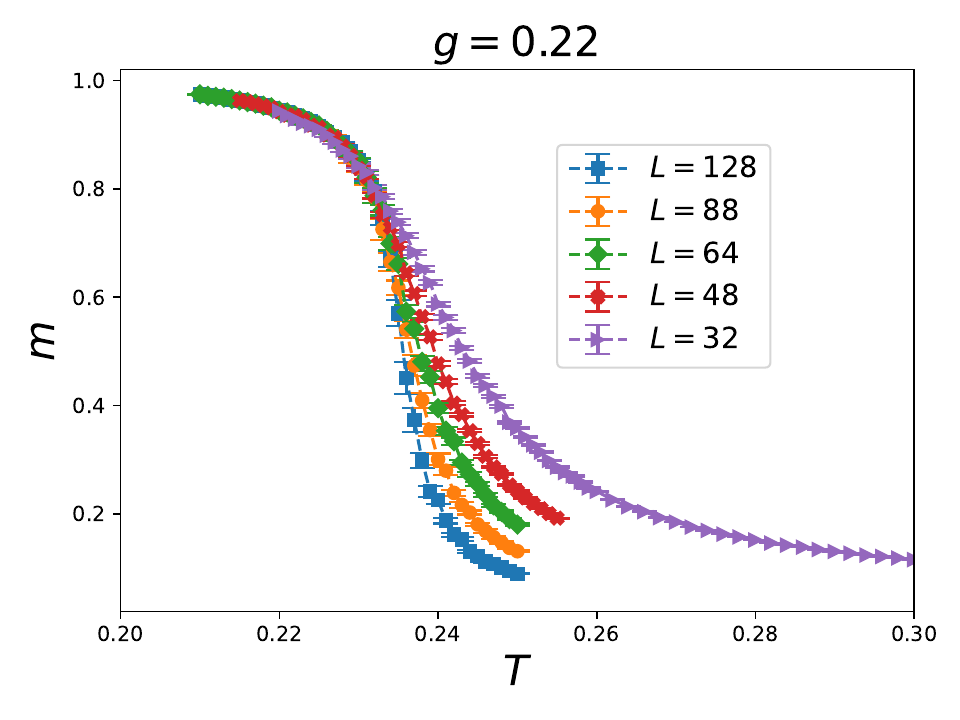}
	}
\end{center}
\caption{$m$ as functions of $T$ for several $L$ for $g=0.20$ (left) and $g = 0.22$ (right).}
\label{m1}
\end{figure}

\begin{figure}
\begin{center}
	\hbox{
		\includegraphics[width=0.45\textwidth]{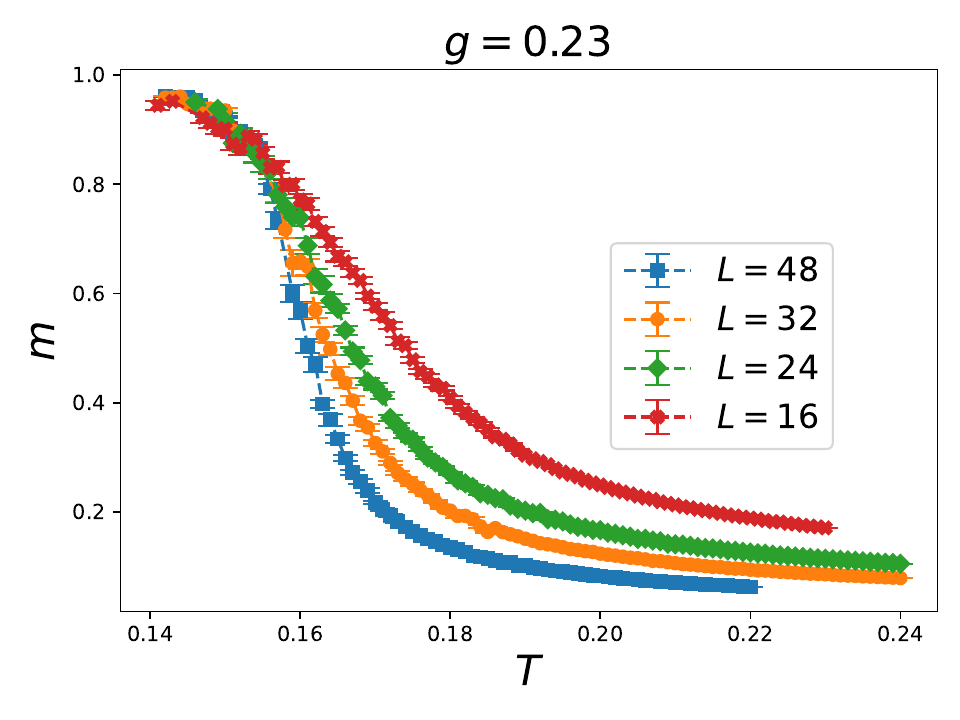}
		\includegraphics[width=0.45\textwidth]{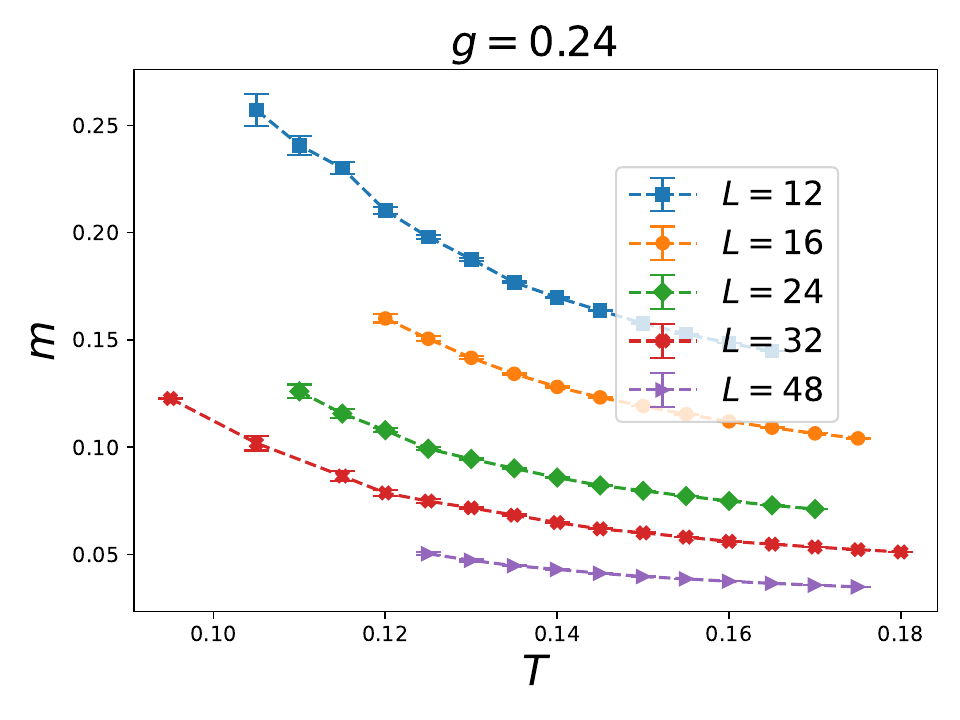}
	}
\end{center}
\caption{$m$ as functions of $T$ for several $L$ for $g=0.23$ (left) and $g = 0.24$ (right).}
\label{m2}
\end{figure}

For $g = 0.20, 0.22, 0.23$ and 0.24, the associated $m$ as functions of $T$ for several $L$ are shown as the left and the right panels of figs.~\ref{m1} and \ref{m2}, respectively. These figs. reveal the message that  
the data of $g=0.24$ are a little bit away from its $T_c$.

If the considered phase transitions are second order and belong to the
2D Ising universality class, then when the data of $mL^{\beta/\nu}$ are
treated as functions of $tL^{1/\nu}$ (Here $t = \left(T-T_c\right)/T_c$), a smooth data collapse curve should emerge
with $\beta = 0.125$ and $\nu = 1.0$.

\begin{figure}
	\begin{center}
		\hbox{
			\includegraphics[width=0.45\textwidth]{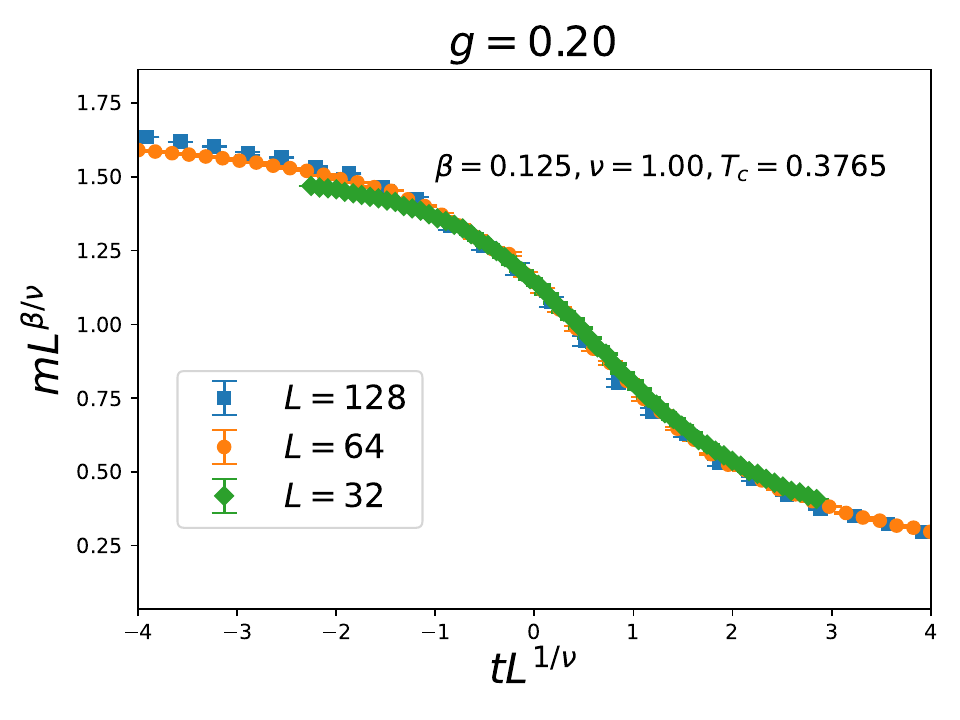}
			\includegraphics[width=0.45\textwidth]{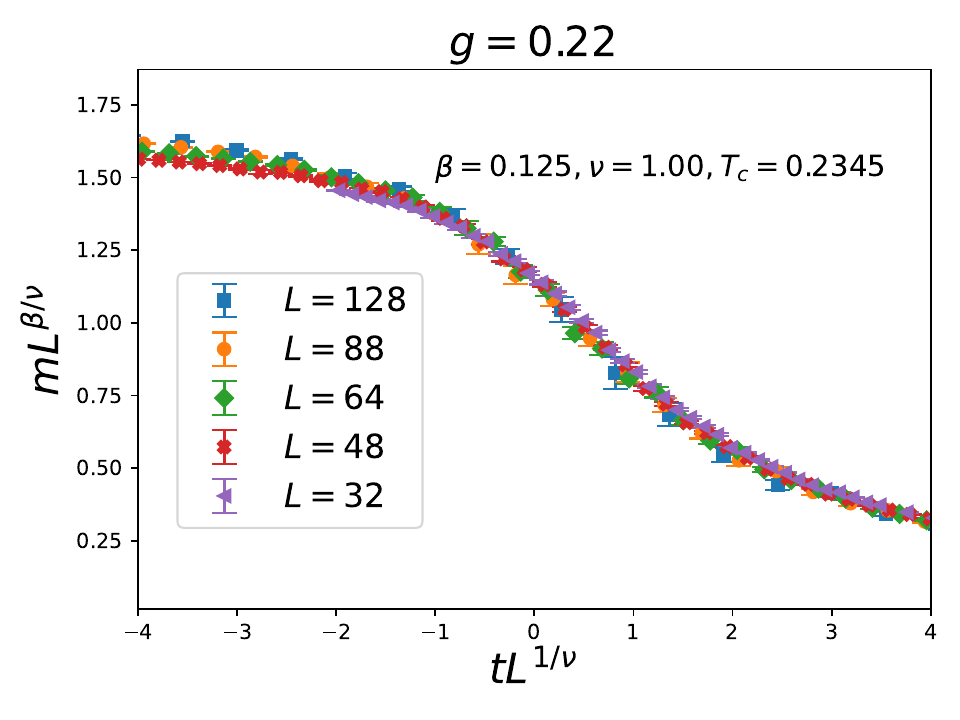}
		}
	\end{center}
	\caption{$mL^{\beta/\nu}$ as functions of $tL^{1/\nu}$ for several $L$ for $g=0.20$ (left) and $g = 0.22$ (right).}
	\label{m3}
\end{figure}

\begin{figure}
	\begin{center}
		\hbox{
			\includegraphics[width=0.45\textwidth]{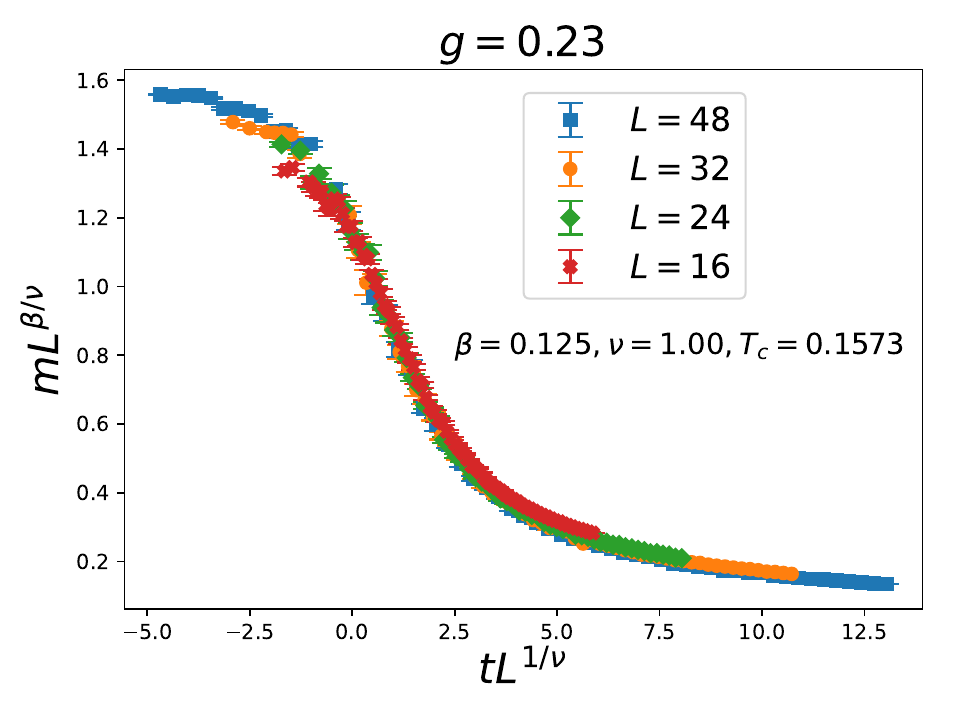}
			\includegraphics[width=0.45\textwidth]{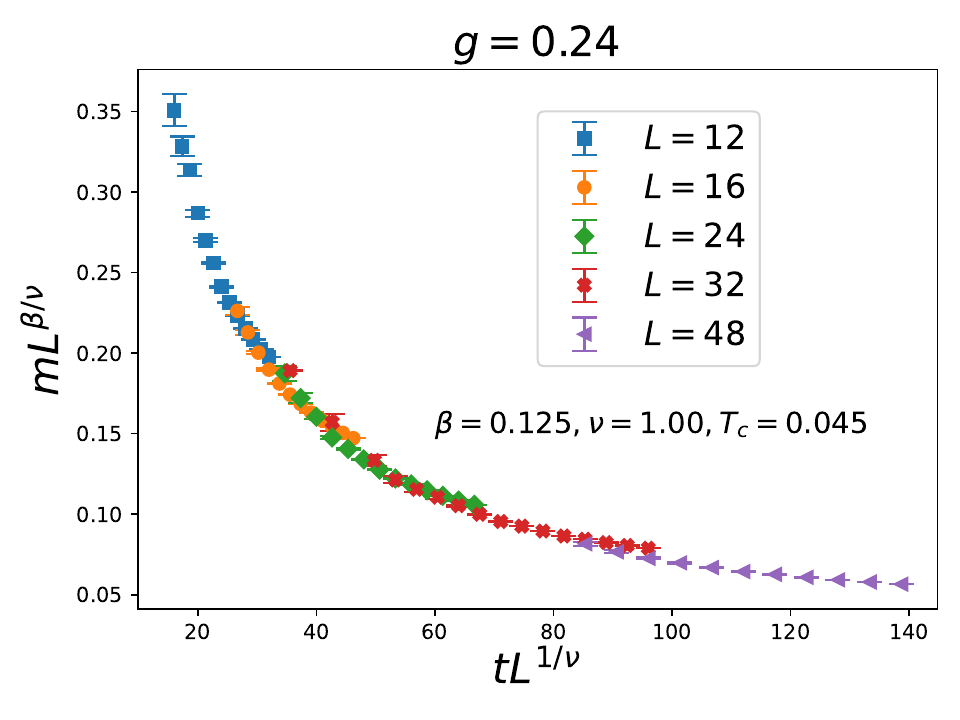}
		}
	\end{center}
	\caption{$mL^{\beta/\nu}$ as functions of $tL^{1/\nu}$ for several $L$ for $g=0.23$ (left) and $g = 0.24$ (right).}
	\label{m4}
\end{figure}

For $g=0.20,0.22$ and 0.23, using the $T$ associated with the peaks of $C_v$ as well as $\beta = 0.125$ and $\nu = 1.0$,
data collapse curves with reasonable smooth quality do show up, see fig.~\ref{m3} and the left panel of fig.~\ref{m4}. These results suggest convincingly that the phase transitions for $g=0.20, 0.22$ and 0.23 are second order and belong to the 2D Ising universality class.

Interestingly, for $g=0.24$, with $T_c \sim 0.045$, $\beta = 0.125$,
and $\nu = 1.0$, a data collapse of $mL^{\beta/\nu}$ v.s $tL^{1/\nu}$
leads to a curve with acceptable smoothness (The right panel of fig.~\ref{m4}). This implies that for $g=0.24$, $T_c \sim 0.045$. In addition, the transition is second order and is governed by the 2D Ising universality class. As we will demonstrate later, $T_c \sim 0.045$ obtained from $m$ also agrees with
the value of $T_c$ calculated from the quantity $\chi$.

\subsubsection{susceptibility $\chi$}

\begin{figure}
	\begin{center}
		\hbox{
			\includegraphics[width=0.45\textwidth]{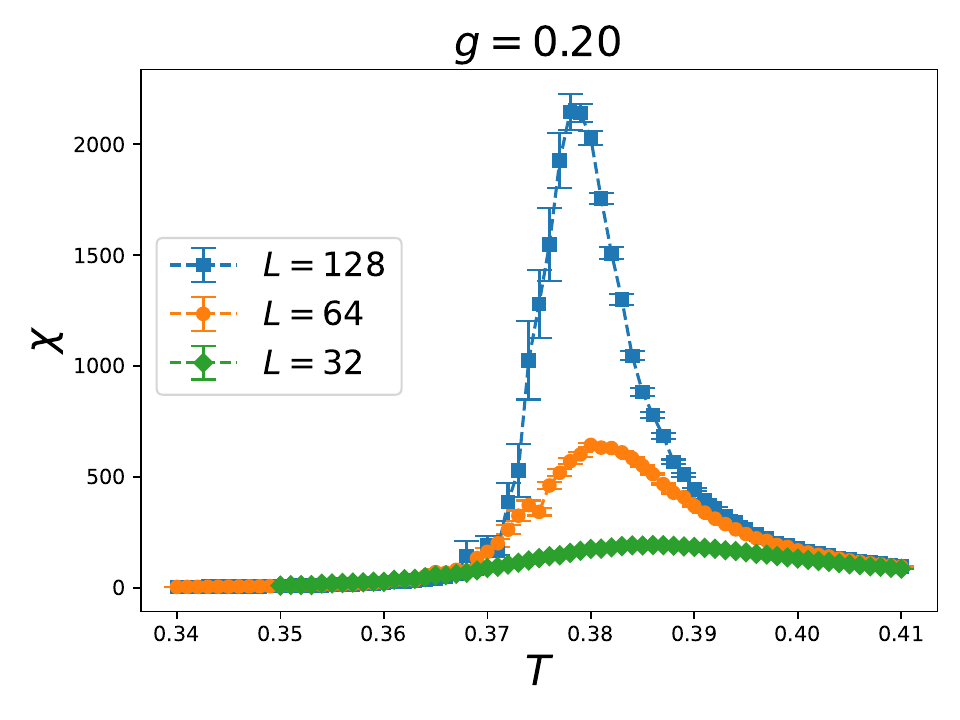}
			\includegraphics[width=0.45\textwidth]{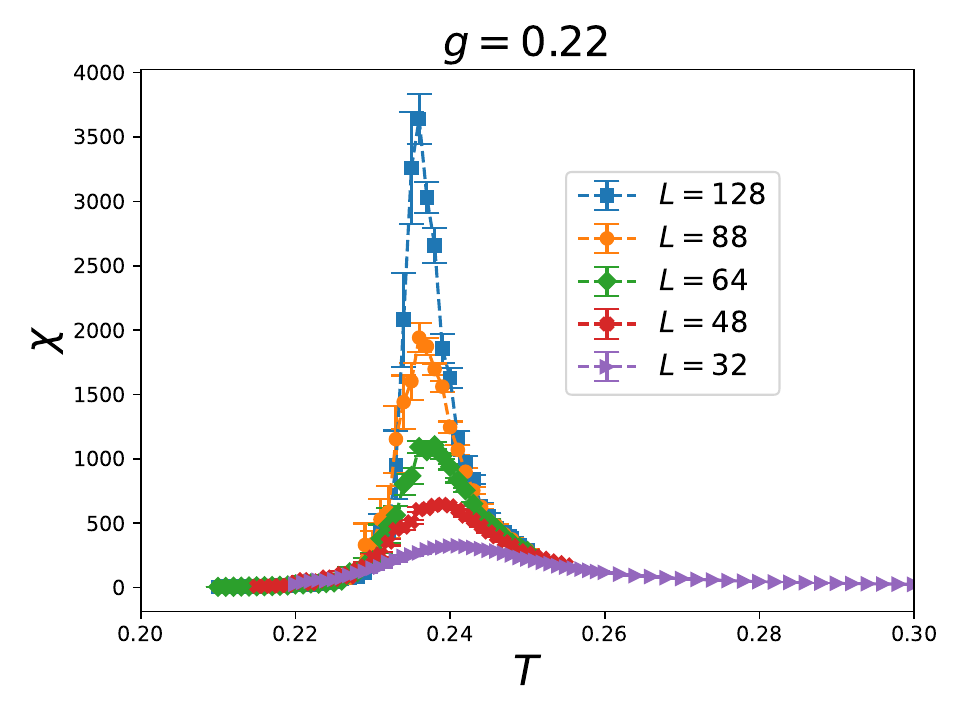}
		}
	\end{center}
	\caption{$\chi$ as functions of $T$ for several $L$ for $g=0.20$ (left) and $g = 0.22$ (right).}
	\label{chi1}
\end{figure}

\begin{figure}
	\begin{center}
		\hbox{
			\includegraphics[width=0.45\textwidth]{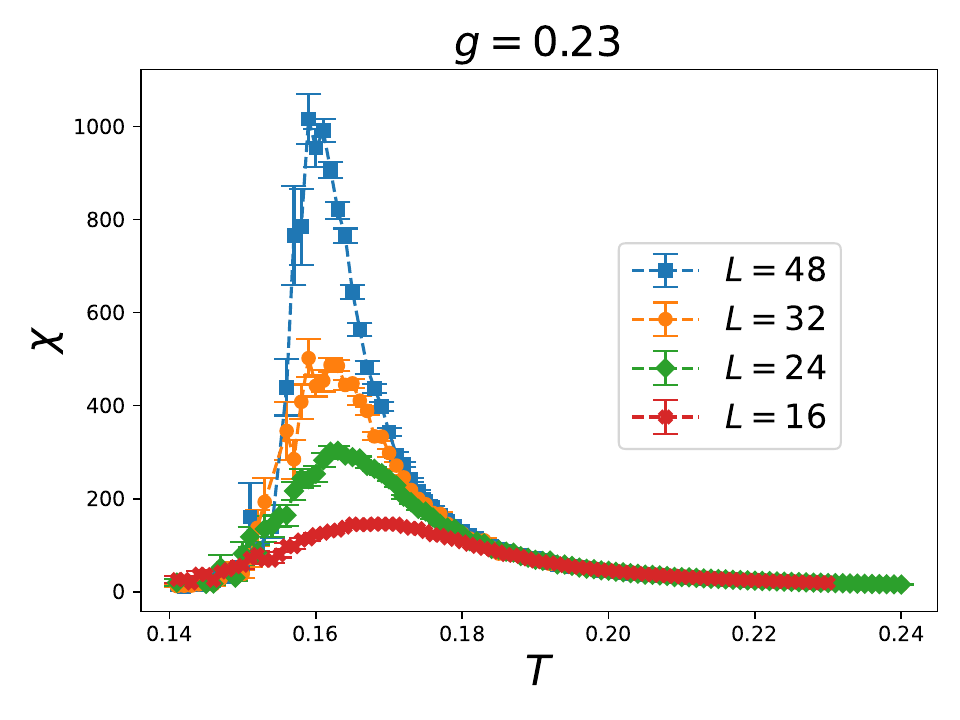}
			\includegraphics[width=0.45\textwidth]{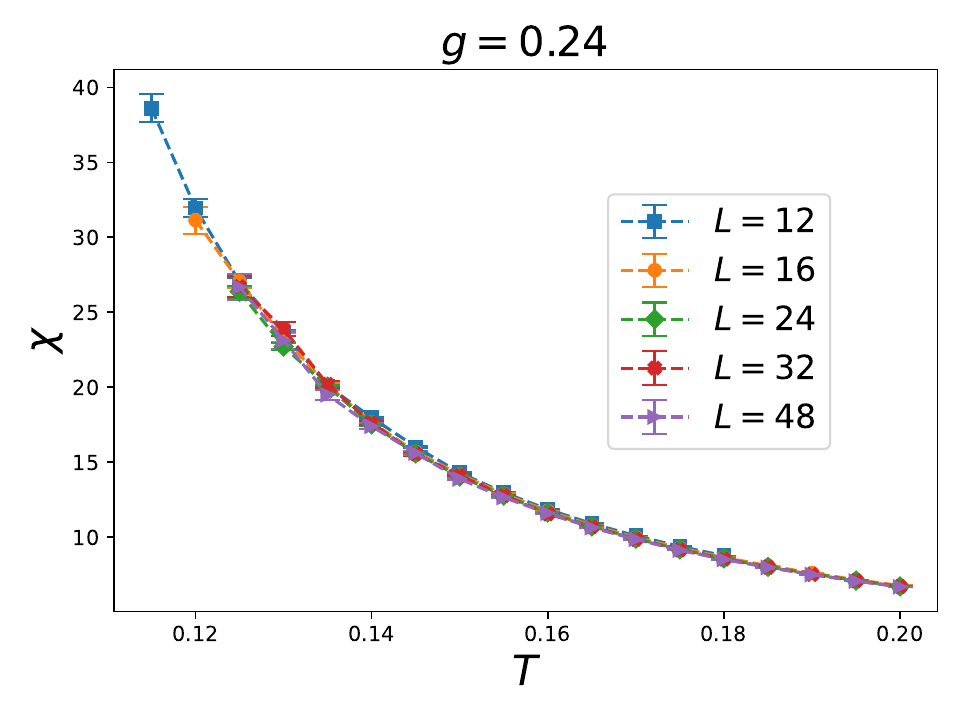}
		}
	\end{center}
	\caption{$\chi$ as functions of $T$ for several $L$ for $g=0.23$ (left) and $g = 0.24$ (right).}
	\label{chi2}
\end{figure}

For $g = 0.20, 0.22, 0.23$ and 0.24, the associated $\chi$ as functions of $T$ for several $L$ are shown as the left and the right panels of figs.~\ref{chi1} and \ref{chi2}, respectively. These figs. reveal the message that  
the data of $g=0.24$ are a little bit away from its $T_c$.

If the considered phase transitions are second order and belong to the
2D Ising universality class, then when the data of $\chi L^{-\gamma/\nu}$ are
treated as functions of $tL^{1/\nu}$, a smooth data collapse curve should emerge
with $\gamma = 1.75$ and $\nu = 1.0$.

\begin{figure}
	\begin{center}
		\hbox{
			\includegraphics[width=0.45\textwidth]{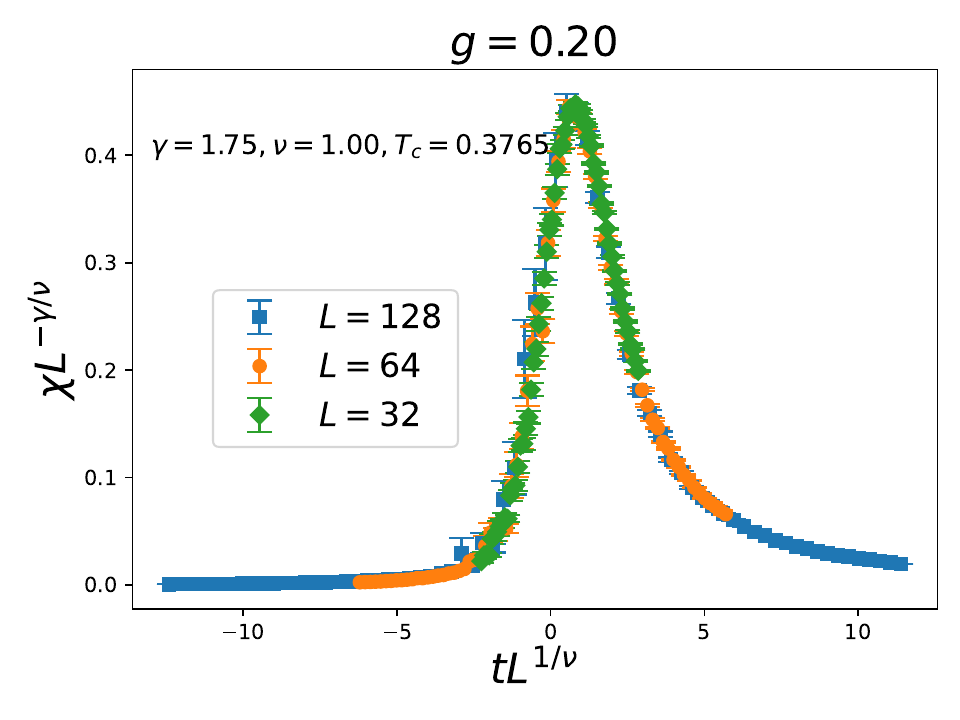}
			\includegraphics[width=0.45\textwidth]{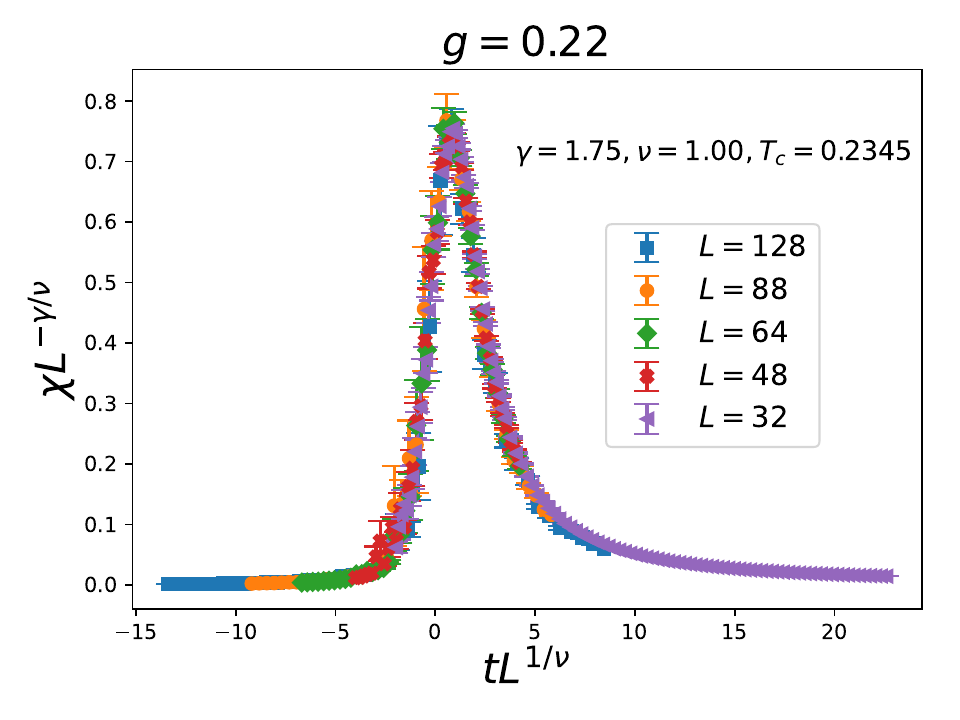}
		}
	\end{center}
	\caption{$\chi L^{-\gamma/\nu}$ as functions of $tL^{1/\nu}$ for several $L$ for $g=0.20$ (left) and $g = 0.22$ (right).}
	\label{chi3}
\end{figure}

\begin{figure}
	\begin{center}
		\hbox{
			\includegraphics[width=0.45\textwidth]{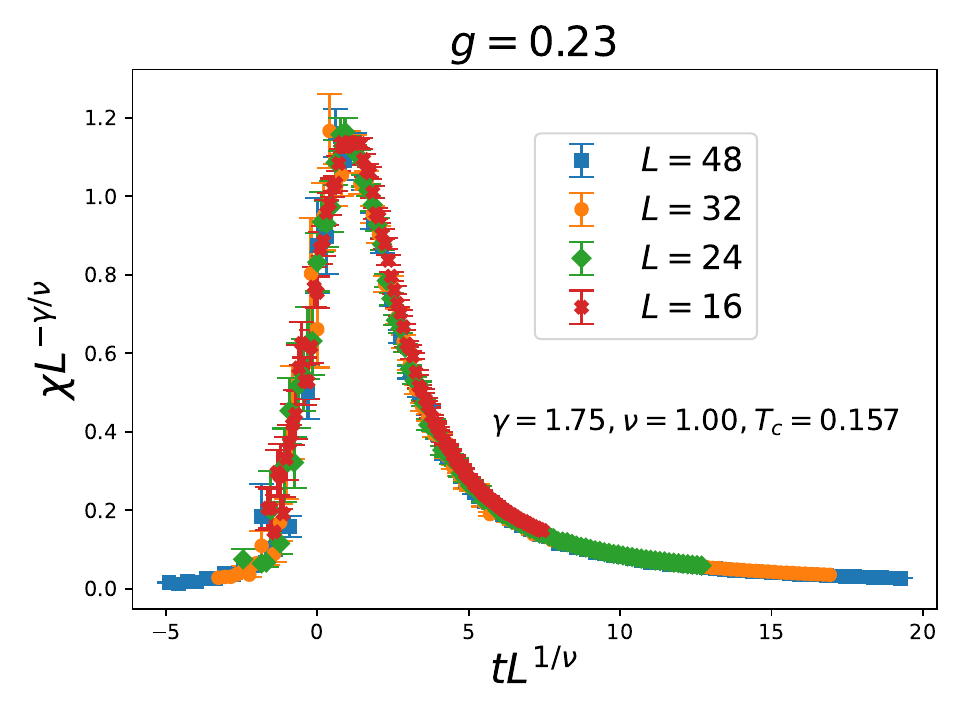}
			\includegraphics[width=0.45\textwidth]{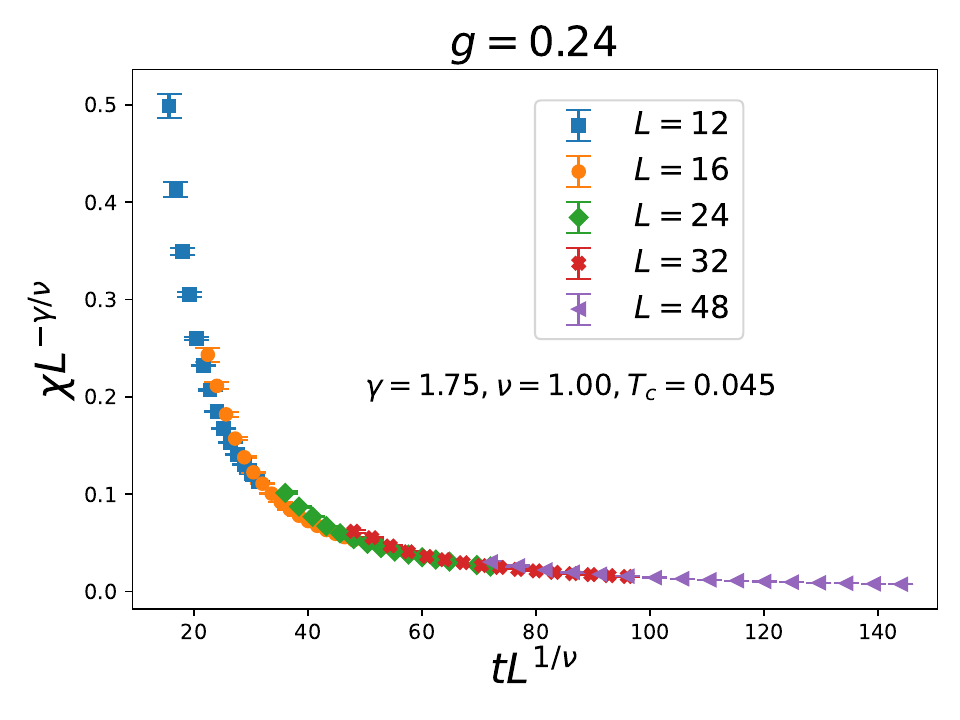}
		}
	\end{center}
	\caption{$\chi L^{-\gamma/\nu}$ as functions of $tL^{1/\nu}$ for several $L$ for $g=0.23$ (left) and $g = 0.24$ (right).}
	\label{chi4}
\end{figure}

For $g=0.20,0.22$ and 0.23, using the $T$ around the peaks of $C_v$ as well as $\gamma = 1.75$ and $\nu = 1.0$,
data collapse curves with reasonable smooth quality do show up, see fig.~\ref{chi3} and the left panel of fig.~\ref{chi4}. These results suggest convincingly that the phase transitions for $g=0.20, 0.22$ and 0.23 are second order and belong to the 2D Ising universality class.

Interestingly, for $g=0.24$, with $T_c \sim 0.045$, $\gamma = 1.75$,
and $\nu = 1.0$, a data collapse of $\chi L^{-\gamma/\nu}$ v.s $tL^{1/\nu}$
leads to a curve with acceptable smoothness (The right panel of fig.~\ref{chi4}). This implies that
for $g=0.24$, $T_c \sim 0.045$. Moreover, the transition is second order and is governed by the 2D Ising universality class as well. The outcome of $T_c \sim 0.045$ obtained from $\chi$ agrees with
the value of $T_c$ calculated from the quantity $m$ previously.

One may argue that the appearance of single data collapse curve in 
the right panel of fig.~\ref{chi4} is due to the fact that the original data without conducting the scaling procedure are already lie on a single curve, i.e. the right panel of fig.~\ref{chi2}.

\begin{figure}
	\begin{center}
		\hbox{
			\includegraphics[width=0.45\textwidth]{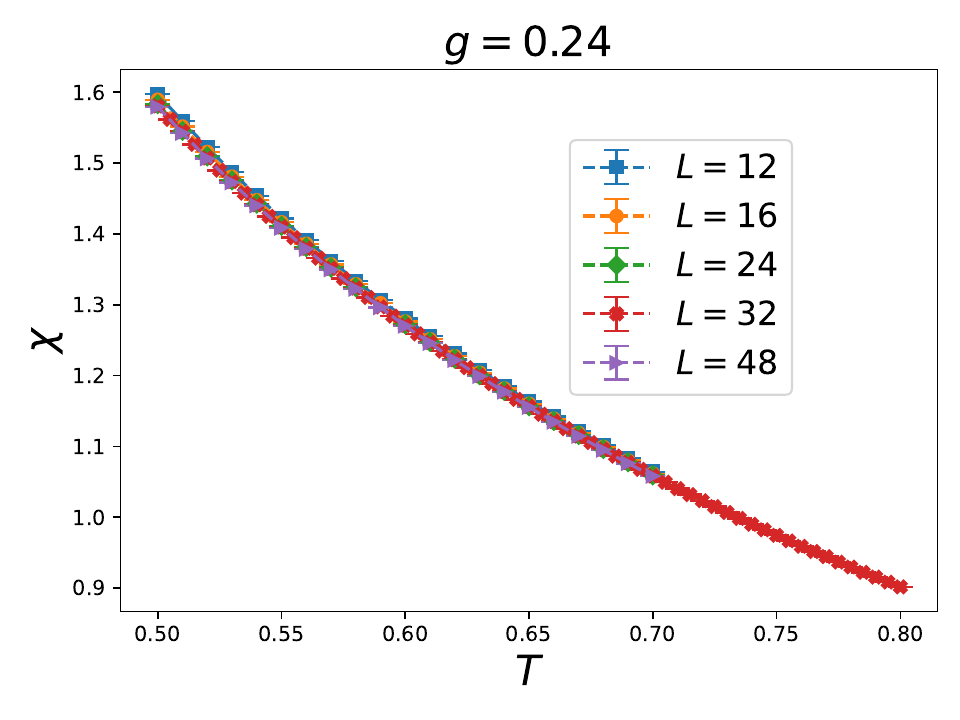}
			\includegraphics[width=0.45\textwidth]{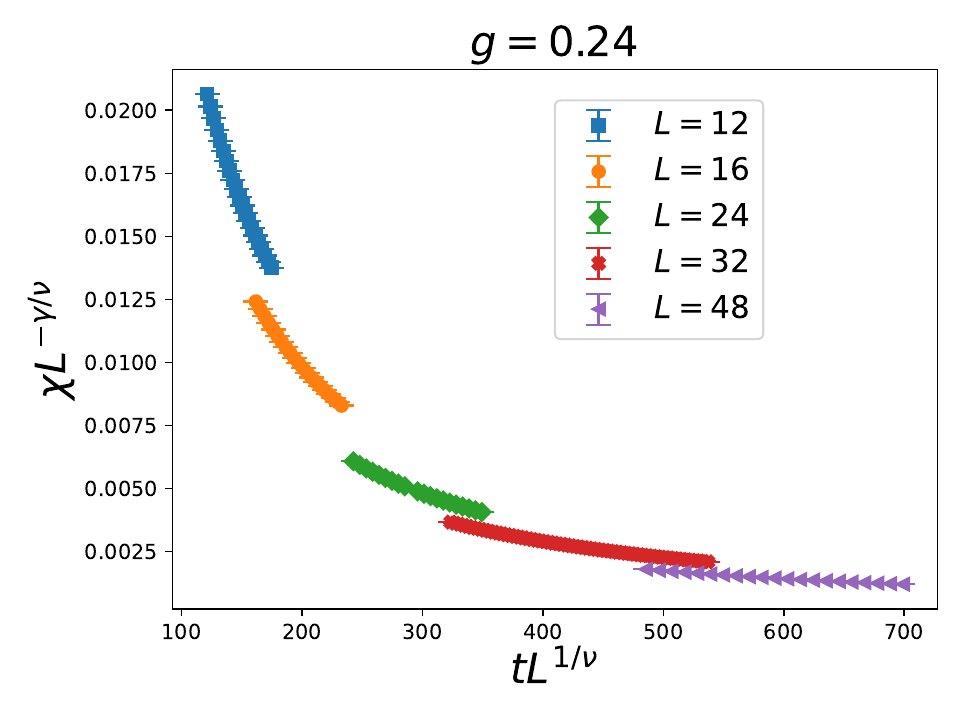}
		}
	\end{center}
	\caption{$\chi $ as functions of $T$ for several $L$ for $g=0.24$ (left). $\chi L^{-\gamma/\nu}$ as functions of $tL^{1/\nu}$ for several $L$ for $g=0.24$ (right).}
	\label{chi5}
\end{figure}

We would like to point out that the left panel of fig.~\ref{chi5} demonstrate $\chi$ as functions of $T$ for several $L$ at the temperature region of 0.5 to 0.8. The figure shows the data of various
$L$ fall into a single curve. After performing the scaling, the associated result is depicted in the right panel of fig.~\ref{chi5}, and smooth data collapse curve is not observed. The figure is obtained with $\gamma = 1.75$, $\nu = 1.0$, and $T_c = 0.045$. Therefore the single data collapse curve shown up in the right panel of fig.~\ref{chi4} is due to the fact that the associated data fulfill the scaling ansatz.

\subsection{Phase transitions for $g > 0.25$}

For $g > 0.25$, it is argued that the ground states are macroscopically degenerate \cite{Bob16,Zuk19}. As a result, no physical quantity can be used to construct the relevant order parameter. Therefore, we rely on the behaviors of the $T$-dependence of $C_v$ and $E$ to determine whether there is any phase transition for a given $g$.

Before presenting the results, we would like to point out that similar to the cases of $g < 0.25$, every data point shown in the following is obtained from the mean of few thousand to a few ten thousand data points (and each of these data points is the averaged result of performing several thousand to (a) few ten thousand MC sweeps). Similar to the situation of $g < 0.25$, only those data that the emcee returns no warning message and their $\tau_{\text{int}}$ are smaller than certain numbers (100 in most of the cases) are accepted and presented.

Finally, inspired the term super-antiferromagnetic phase introduced in Ref.~\cite{Zuk19}, the stagger-like configurations are employed as the initial configurations to start the MC simulations. Since the algorithm employed can only perform local updates and the ground states are macroscopically degenerate, this choice of initial configurations to start the MC simulations may help to reach the true equilibrium steady states, in particular at the low temperature region. 

Indeed, as we will demonstrate later in appendix B, the outcomes associated with the stagger-like configurations are more stable and have shorter integrated autocorrelation times in the low-$T$ region, although it comes with one disadvantage (This will be shown explicitly as well).

\subsubsection{$g=1.0$}

\begin{figure}
	\begin{center}
		\hbox{
			\includegraphics[width=0.45\textwidth]{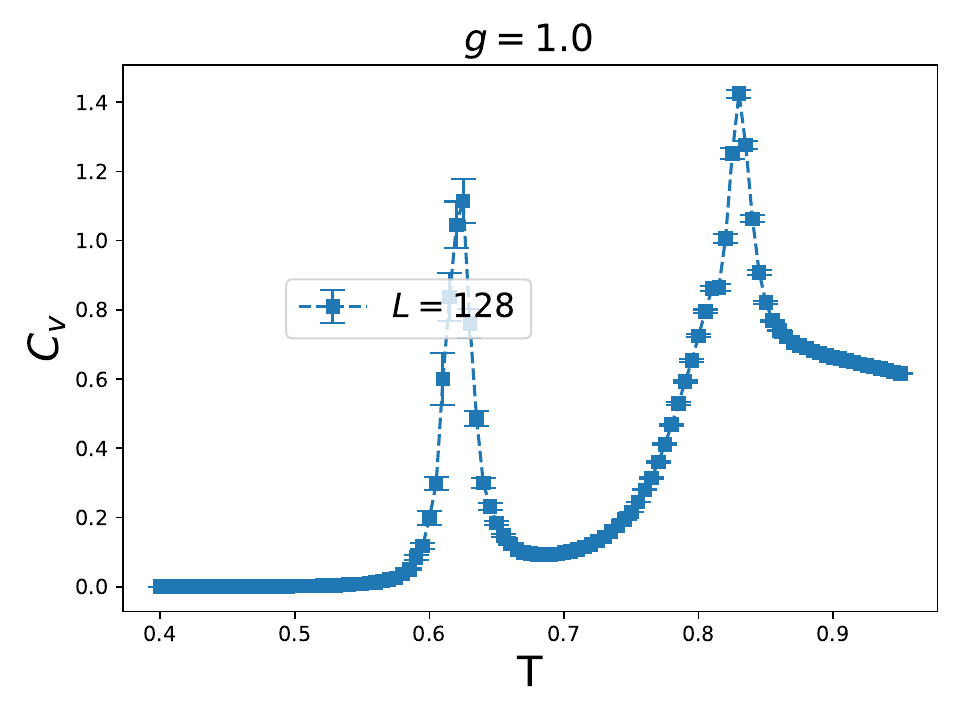}
			\includegraphics[width=0.45\textwidth]{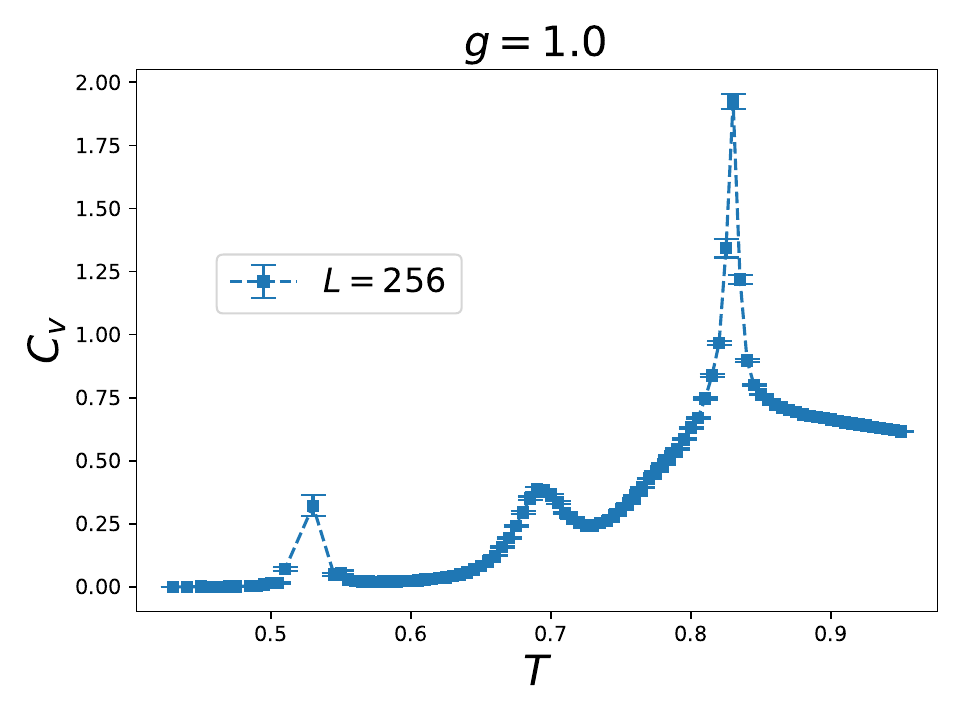}
		}
	\end{center}
	\caption{$C_v$ as functions of $T$ for $L = 128$ (left) and $L = 256$ (right) for $g = 1.0$.}
	\label{cv6}
\end{figure}

\begin{figure}
	\begin{center}
		\hbox{
			\includegraphics[width=0.45\textwidth]{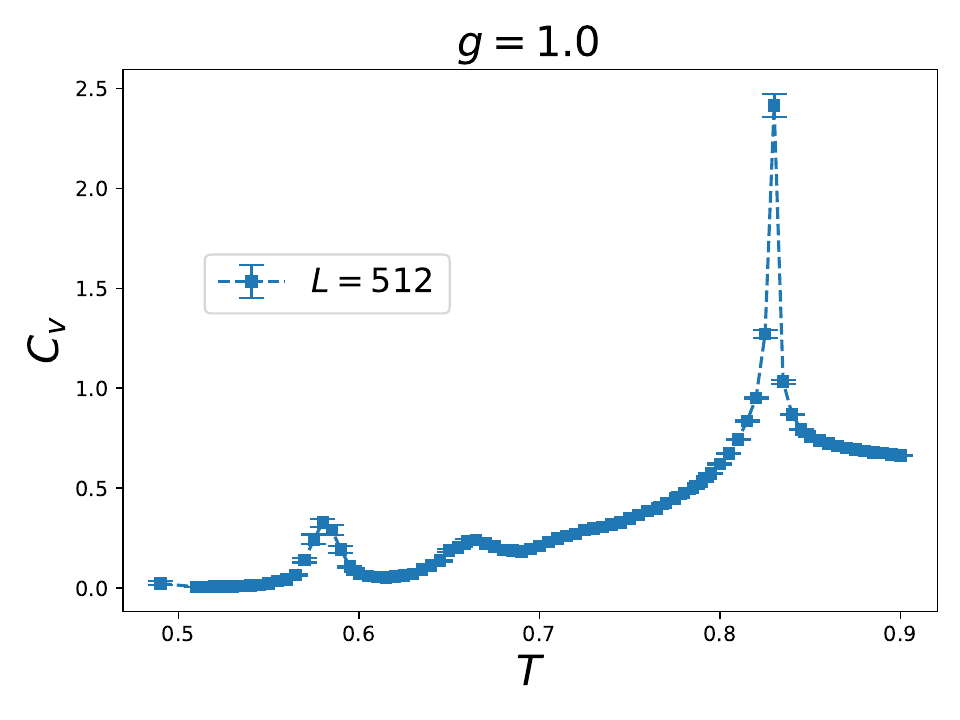}
			\includegraphics[width=0.45\textwidth]{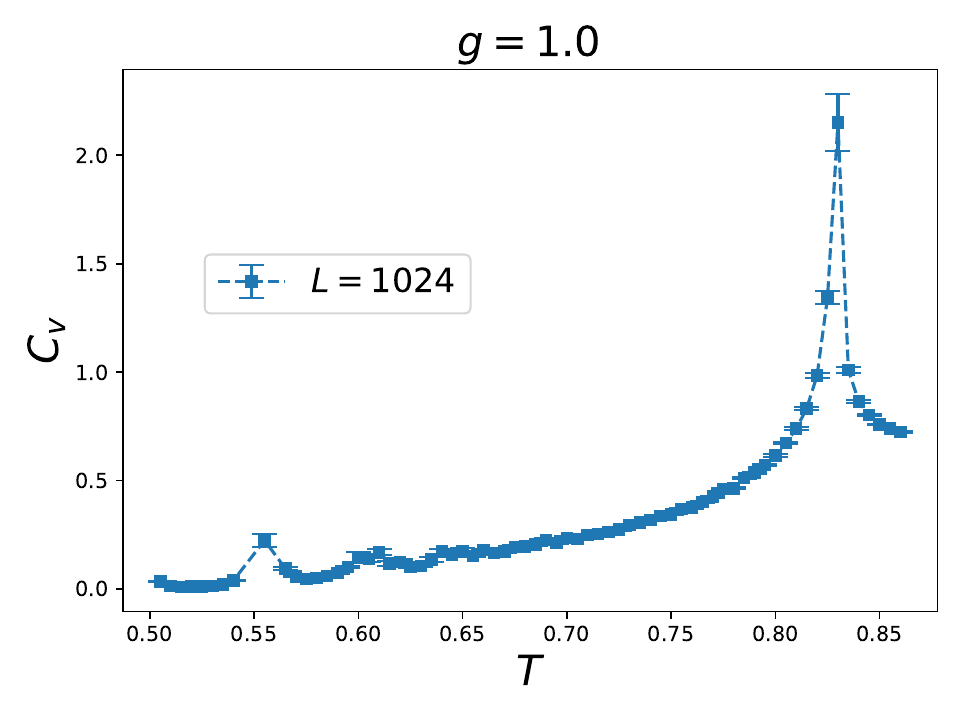}
		}
	\end{center}
	\caption{$C_v$ as functions of $T$ for $L = 512 $ (left) and $L = 1024$ (right) for $g = 1.0$.}
	\label{cv7}
\end{figure}

\begin{figure}
	\begin{center}
			\includegraphics[width=0.5\textwidth]{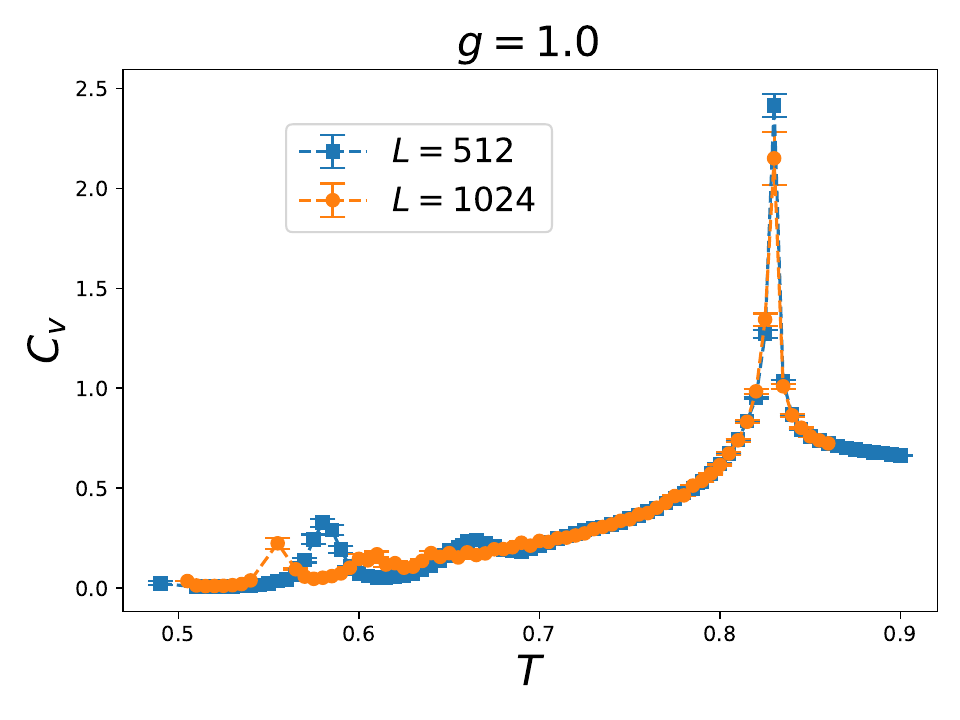}
	\end{center}
	\caption{$C_v$ as functions of $T$ for $L = 512$ and $L = 1024$ for $g = 1.0$.}
	\label{cv8}
\end{figure}

For $g=1.0$, $C_v$ as functions of $T$ for various $L$ are shown in
figs.~\ref{cv6} and \ref{cv7}.
Interestingly, despite the fact that there are multiple peaks in these figures, only the one near $T \sim 0.84$ survives when the linear system size increases from $L=128$ to $L=1024$.
This means the appearance of multiple peaks when $C_v$ is considered as a function of $T$ is a finite-size effect. In addition,   
the heights of these surviving peaks saturate to a value around 2.25, see fig.~\ref{cv8}. 

\begin{figure}
	\begin{center}
		\includegraphics[width=0.5\textwidth]{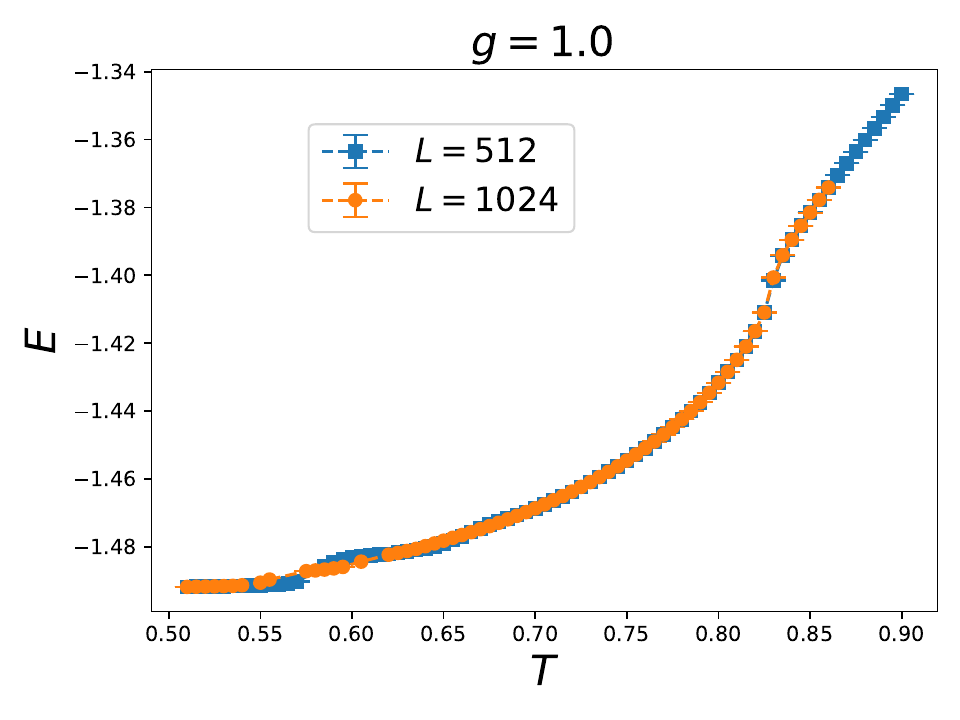}
	\end{center}
	\caption{$E$ as functions of $T$ for $L = 512$ and $L = 1024$ for $g = 1.0$.}
	\label{E1}
\end{figure}

$E$ as functions of $T$ for $L=512$ and $L=1024$ are demonstrated in fig.~\ref{E1}. The figure indicates that there is an apparent jump
near $T \sim 0.83$. If the phase transition is first order, then one expected $C_v$ scales
with $L^2$. Hence,
the results presented in figs.~\ref{cv8} and \ref{E1} support the scenario that there is a second phase transition with $T_c \sim 0.83$ for $g = 1.0$.

\begin{figure}
	\begin{center}
		\includegraphics[width=0.5\textwidth]{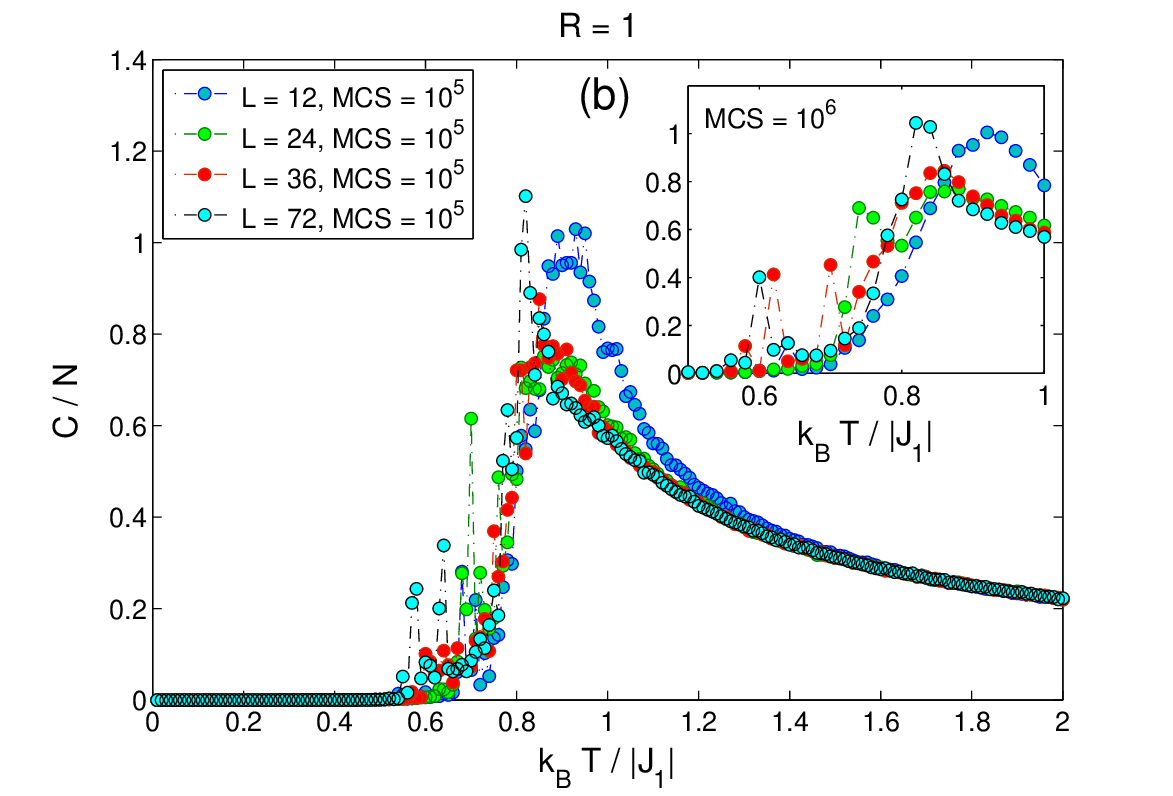}
	\end{center}
	\caption{$C_v$ as functions of $T$ for several $L$ for $g = 1.0$.
	The figure is taken from the arXiv version of Ref.~\cite{Zuk19}. The $R$ appearing in the title of this figure is the $g$ used here.}
	\label{cv9}
\end{figure}

Fig.~\ref{cv9} is from the arXiv version of Ref.~\cite{Zuk19}, and is the $C_v$ as functions
of $T$ for several $L$ for $g=1.0$. Without doubt, a peak close to $T\sim 0.8$ shows up in the figure. This is consistent with our finding. 
It should be pointed out that the authors of Ref.~\cite{Zuk19} claim
that the phase transition for $g=1.0$ is first order due to a very tiny jump in $E$ around $T=0.6$. Since the peaks of $C_v$ on finite lattices $L$ should scale with $L^2$ for a first-order phase transition, we find that the claim of Ref.~\cite{Zuk19} is unlikely. In particular, our analysis indicates
that $\tau_{\text{int}}$ around $T=0.6$ is (at least) $10^5$ when $L=128$ (Each data point for $L=128$ shown above is the average of performing $10^4$ MC sweeps). Since
the data used in Ref.~\cite{Zuk19} contain at most $10^6$ consecutive 
Monte Carlo outcomes,
it is of high possibility that the results presented in Ref.~\cite{Zuk19} is biased.

\begin{figure}
	\begin{center}
		\includegraphics[width=0.5\textwidth]{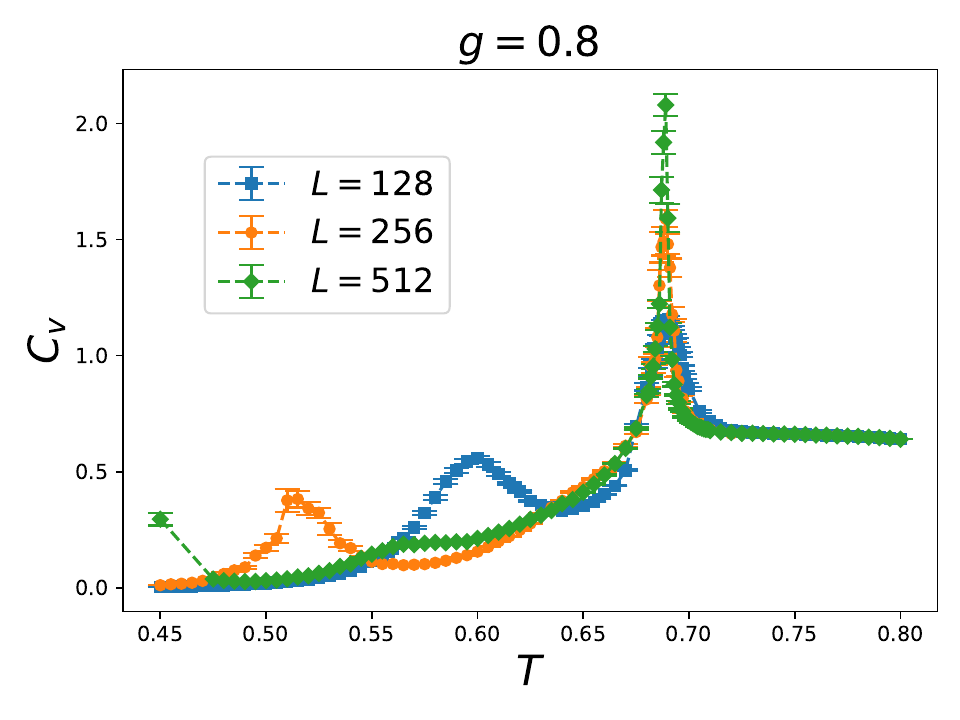}
	\end{center}
	\caption{$C_v$ as functions of $T$ for various $L$ for $g = 0.8$.}
	\label{cv10}
\end{figure}

\begin{figure}
	\begin{center}
		\includegraphics[width=0.5\textwidth]{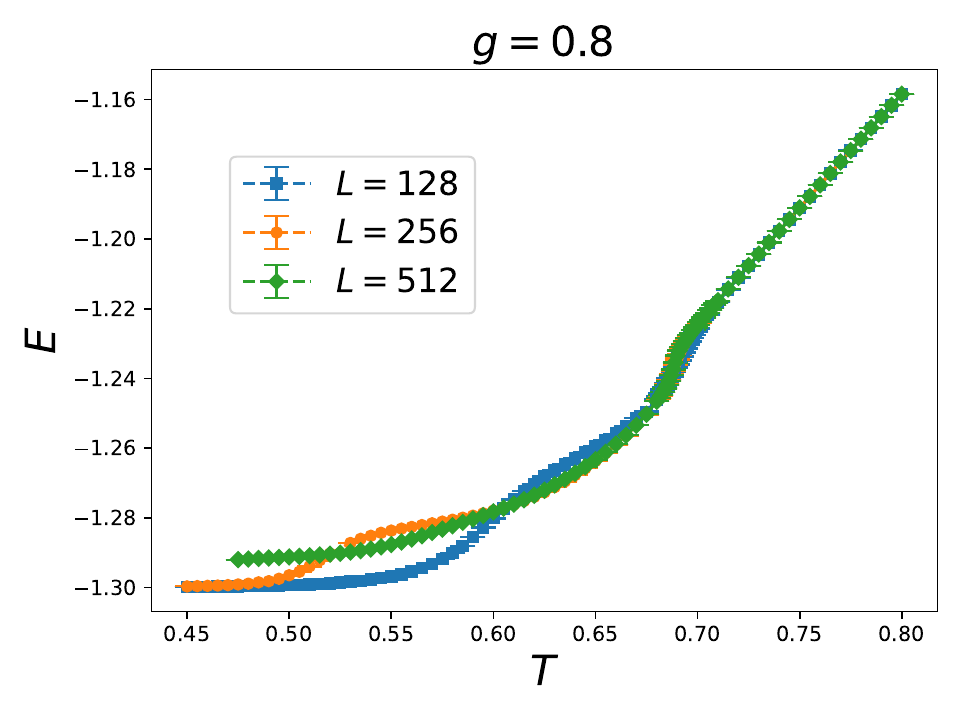}
	\end{center}
	\caption{$E$ as functions of $T$ for various $L$ for $g = 0.8$.}
	\label{E2}
\end{figure}

\begin{figure}
	\begin{center}
		\hbox{
			\includegraphics[width=0.45\textwidth]{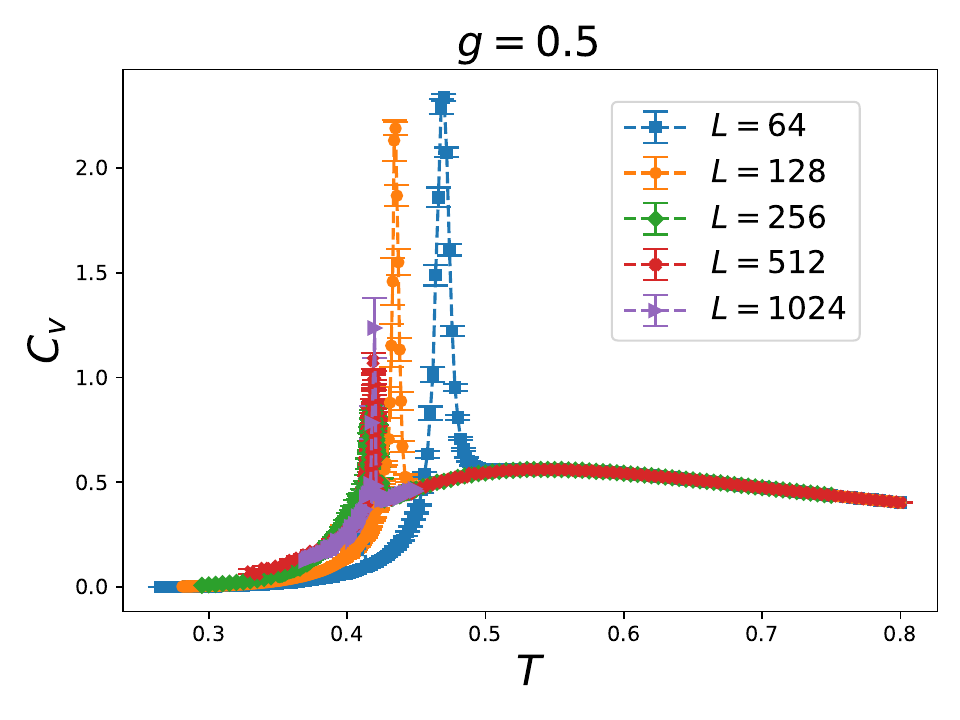}
			\includegraphics[width=0.45\textwidth]{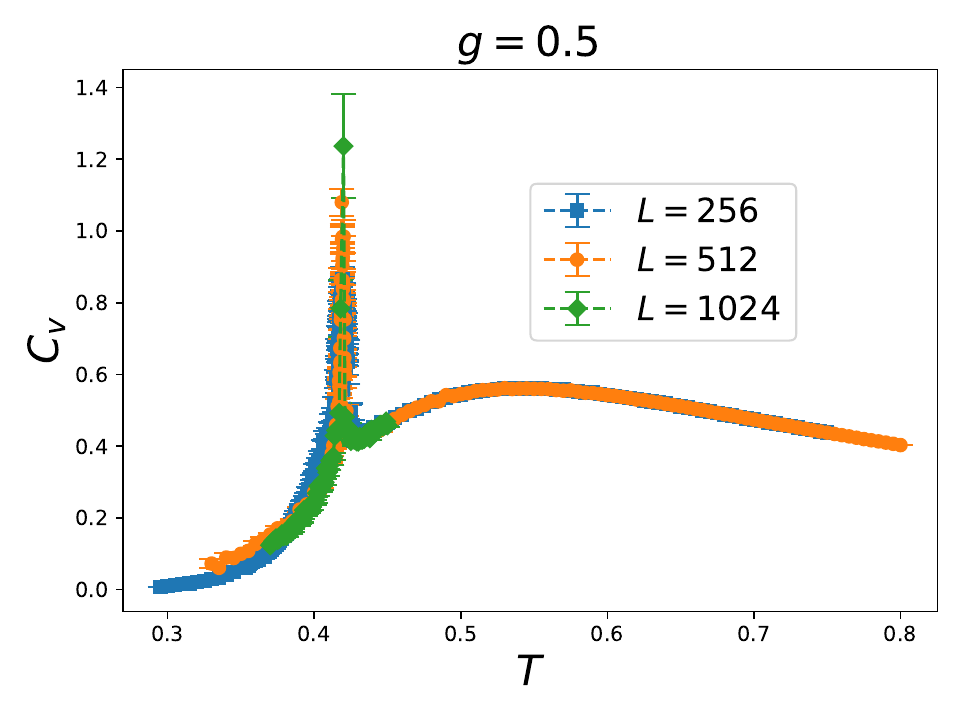}
		}
	\end{center}
	\caption{$C_v$ as functions of $T$ for various $L$ for $g = 0.5$.}
	\label{cv11}
\end{figure}

\subsubsection{$g=0.8$}

\begin{figure}
	\begin{center}
		\includegraphics[width=0.5\textwidth]{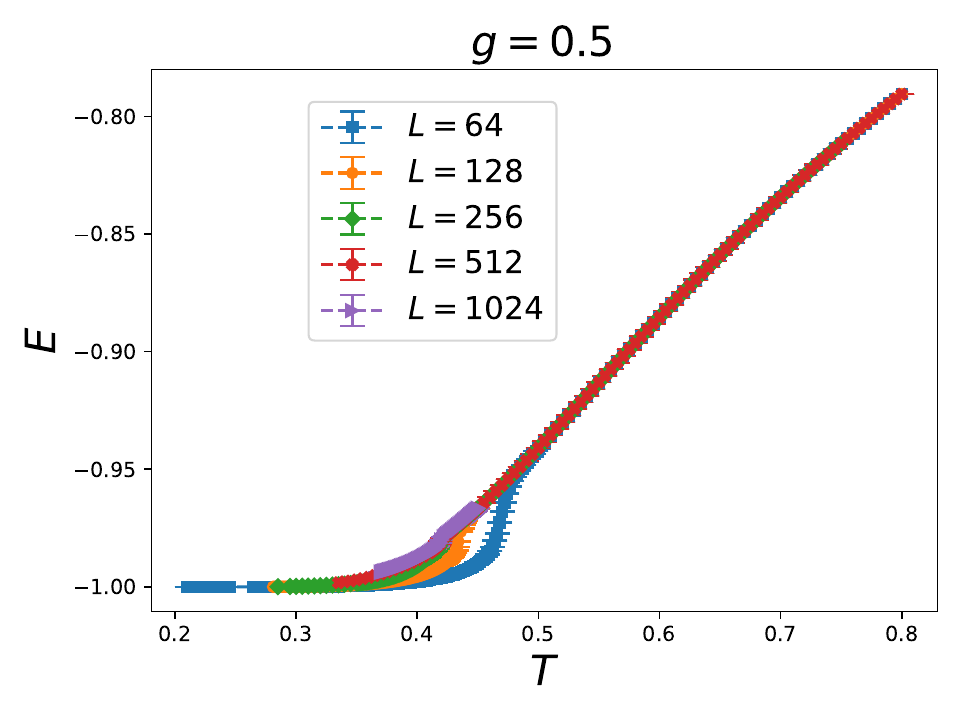}
	\end{center}
	\caption{$E$ as functions of $T$ for various $L$ for $g = 0.5$.}
	\label{E3}
\end{figure}

\begin{figure}
	\begin{center}
		\includegraphics[width=0.5\textwidth]{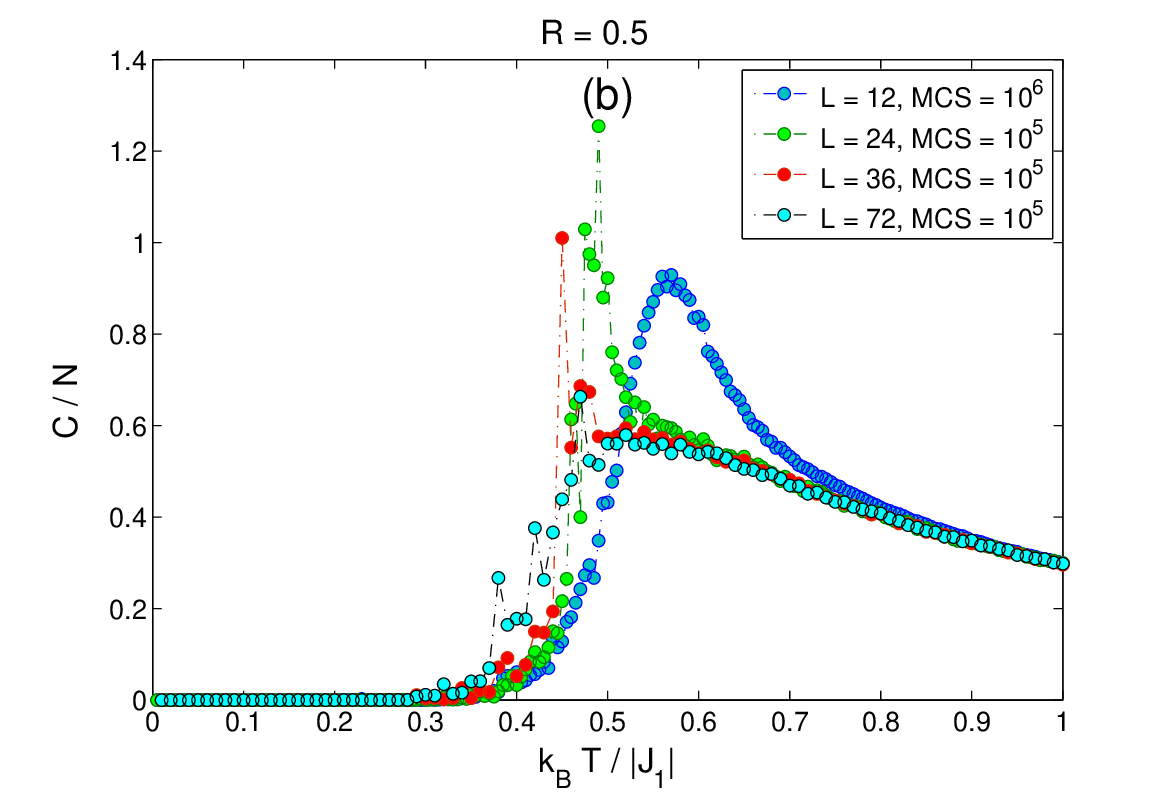}
	\end{center}
	\caption{$C_v$ as functions of $T$ for various $L$ for $g = 0.5$. The figure is taken from the arXiv version of Ref.~\cite{Zuk19}.}
	\label{cv12}
\end{figure}

For $g=0.8$, $C_v$ as functions of $T$ for various $L$ are shown in
fig.~\ref{cv10}.
Similar to the scenario observed for $g=1.0$, multiple peaks show up and only the one around $T=0.69$ survives when $L$ increases from $L=128$ to $L=512$. In addition, since the peak near $T=0.69$ does not
scale with $L^2$, there is also a second order phase transition for $g=0.8$ with $T_c \sim 0.69$.
It should be pointed out that at $T = 0.45$ the magnitude of $C_v$
is larger than those with $T \ge 0.47$. It is not clear at the moment 
the cause for this. Due to the extremely long integrated auto-correlation times, we are not able to obtain results with $T < 0.45$.  

$E$ as functions of $T$ for several $L$ for $g=0.8$ can be depicted in
fig.~\ref{E2}. For $L=512$, a noticeable jump in $E$ occurs near $T\sim 0.67$. This is consistent with that obtained from the quantity $C_v$.

\subsubsection{$g=0.5$}

For $g=0.5$, $C_v$ as functions of $T$ for various $L$ are shown in
fig.~\ref{cv11}. Unlike the cases of $g=1.0$ and $g=0.8$, only one peak appears for each considered $L$. The peak approaches $T \sim 0.41$
as the magnitude of $L$ increases. In addition, the left panel of fig.~\ref{cv11} shows that the height of peak does increase with $L \ge 256$. In other words, the peak unlikely disappears with $L$.
In addition, the heights of the peaks do not scale as $L^2$.
This implies there is a second-order phase transition for $g=0.5$ with $T_c \sim 0.41$.

$E$ as functions of $T$ for various $L$ are demonstrated in fig.~\ref{E3}. The message reveals from the figure agrees with that from $C_v$. In particular, $T_c$ is around 0.41. 
In Ref.~\cite{Zuk19}, for $L=72$, the $C_v$ has a peak around $T\sim 0.46$, see fig.~\ref{cv12}. This agrees reasonably well with our $C_v$ result of $L=64$.

\subsubsection{$g=0.3$}

\begin{figure}
	\begin{center}
		\hbox{
			\includegraphics[width=0.45\textwidth]{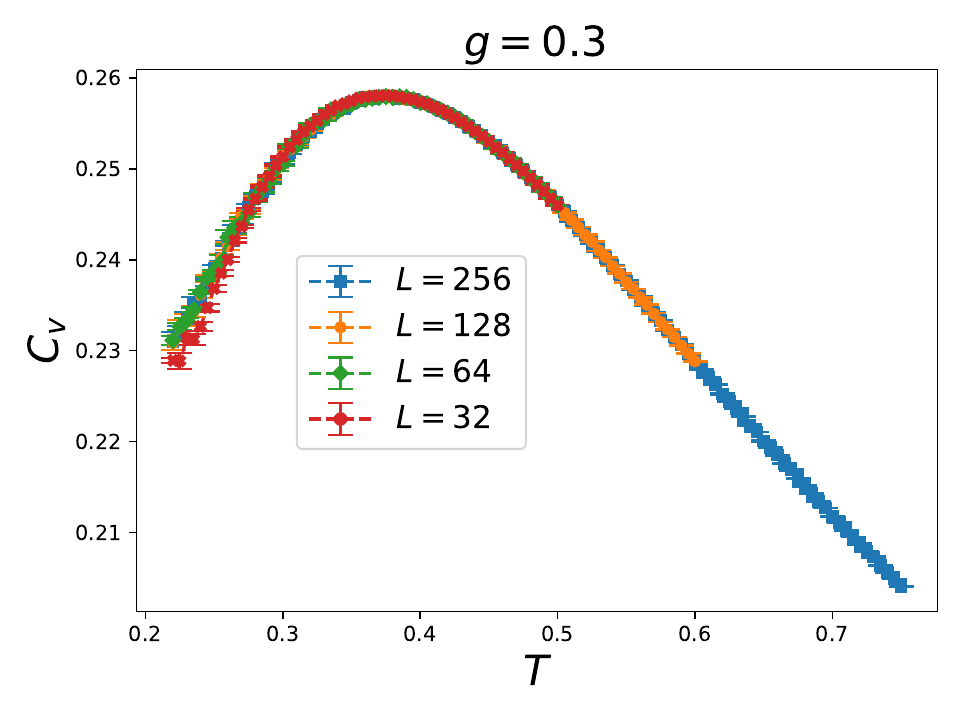}
			\includegraphics[width=0.45\textwidth]{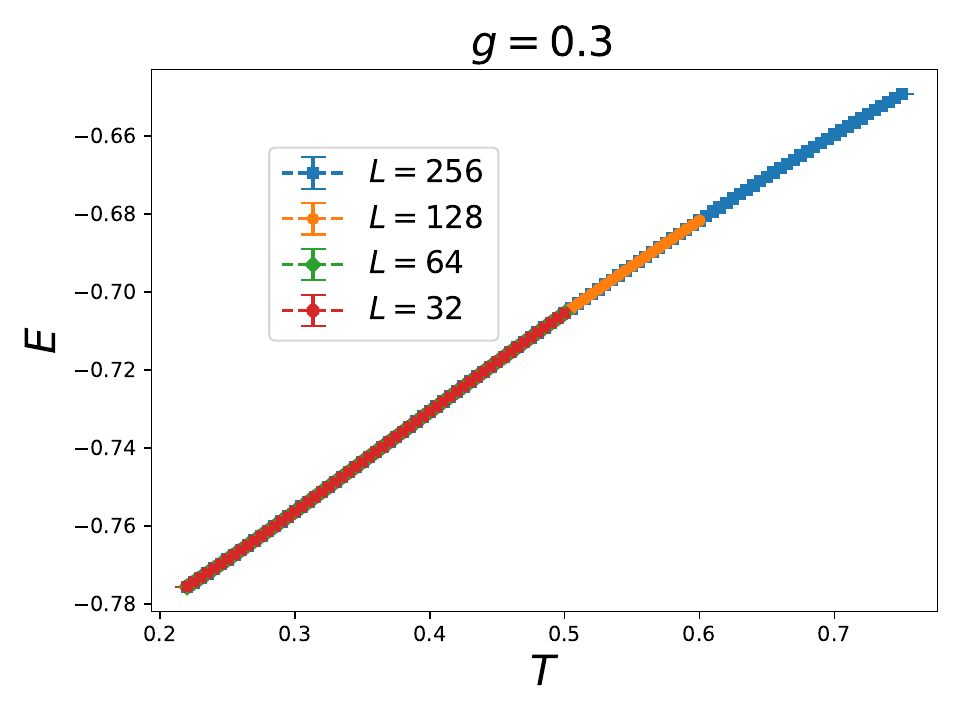}
		}
	\end{center}
	\caption{$C_v$ (left) and $E$ (right) as functions of $T \ge 0.22$ for various $L$ for $g = 0.3$.}
	\label{cve1}
\end{figure}

For $g=0.3$, $C_v$ and $E$ as functions of $T \ge 0.22$ for several $L$
are depicted in the left and the right panels of fig.~\ref{cve1}.
The outcomes of $C_v$ show that a peak appears around $T=0.36$.
However, this peak is round and is not like those of $g=1.0,0.8$ and 0.5 which are much sharper. A abrupt jump in $E$ does not show up
near $T = 0.36$. These results suggest that $T=0.36$ is not a critical
point. If a critical temperature $T_c$ really exists, then one must have $T_c < 0.22$.

\begin{figure}
	\begin{center}
		\hbox{
			\includegraphics[width=0.45\textwidth]{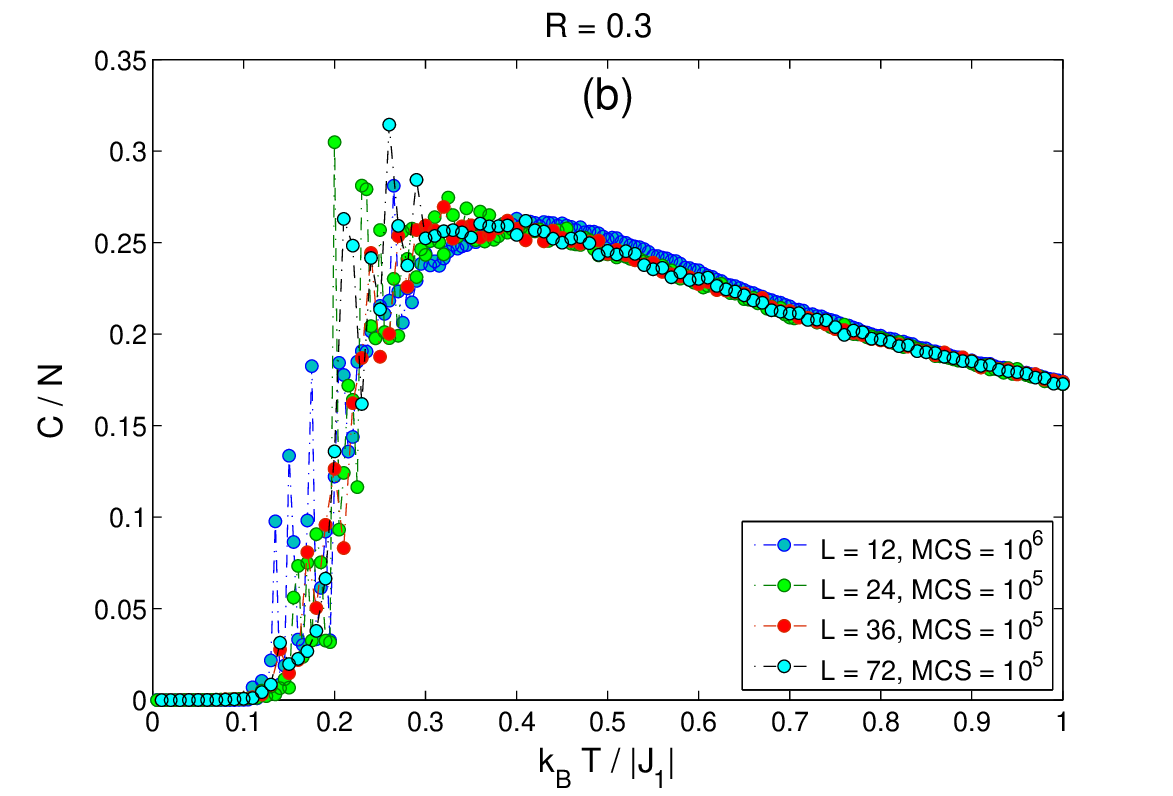}
			\includegraphics[width=0.45\textwidth]{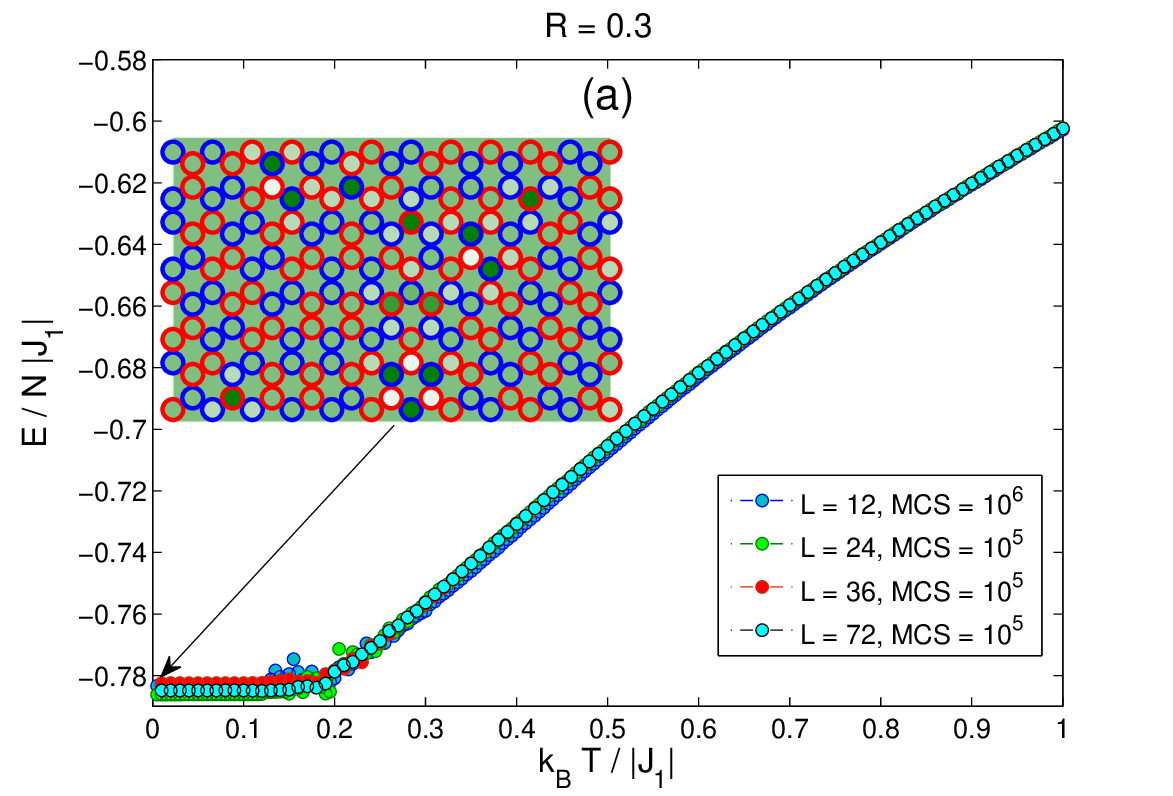}
		}
	\end{center}
	\caption{$C_v$ (left) and $E$ (right) as functions of $T$ for various $L$ for $g = 0.3$. These figures are taken from the arXiv version of Ref.~\cite{Zuk19}.}
\label{CvEZuk19}
\end{figure}

Fig.~\ref{CvEZuk19} shows the $C_v$ and $E$ obtained in Ref.~\cite{Zuk19}. Fig.~\ref{CvEZuk19}
is of high similarity to our results demonstrated in fig.~\ref{cve1}. Indeed, round-shape peaks in $C_v$ around $T\sim 0.36$ appear in both figs (Notice the data in fig.~\ref{CvEZuk19} fluctuate dramatically for $T \le 0.3$).
The heights of the peaks are both around 0.26. 
Finally,
both figs. show no sudden rise in $E$ near $T\sim 0.36$. The agreement mentioned above
again confirms the correctness of our MC codes.   

From figs.~\ref{cve1} and \ref{CvEZuk19}, it is concluded that
there is no phase transition for $g=0.3$ as one moves from low-temperature region to high-temperature region.
The outcomes of $C_v$ with $T \le 0.22$ may be obtained using brute force simulations.

\begin{figure}
	\begin{center}
		\hbox{
			\includegraphics[width=0.33\textwidth]{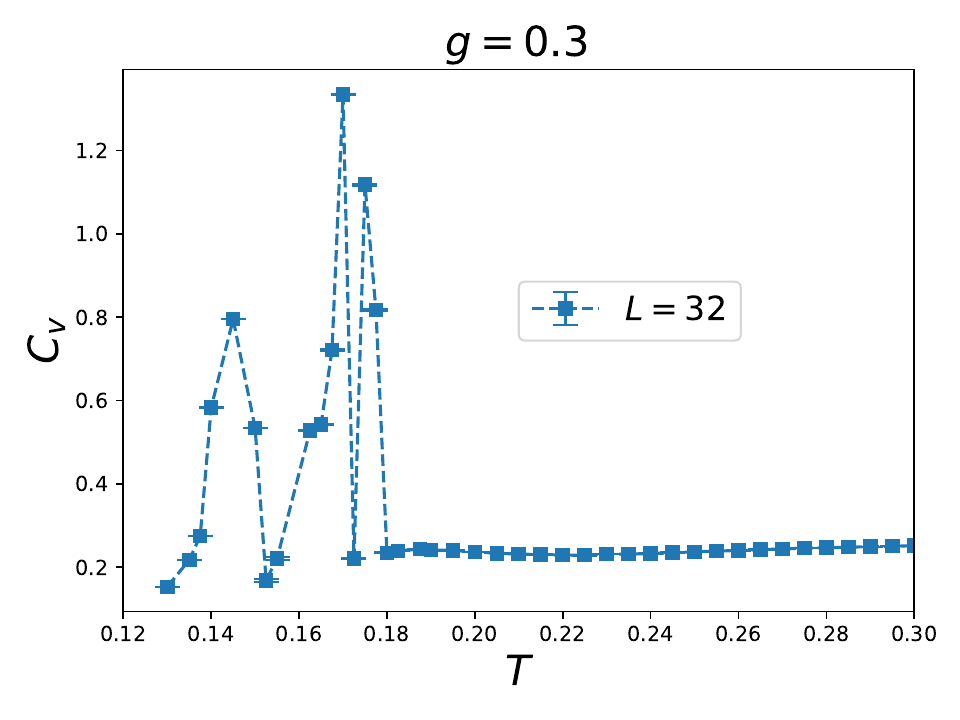}
			\includegraphics[width=0.33\textwidth]{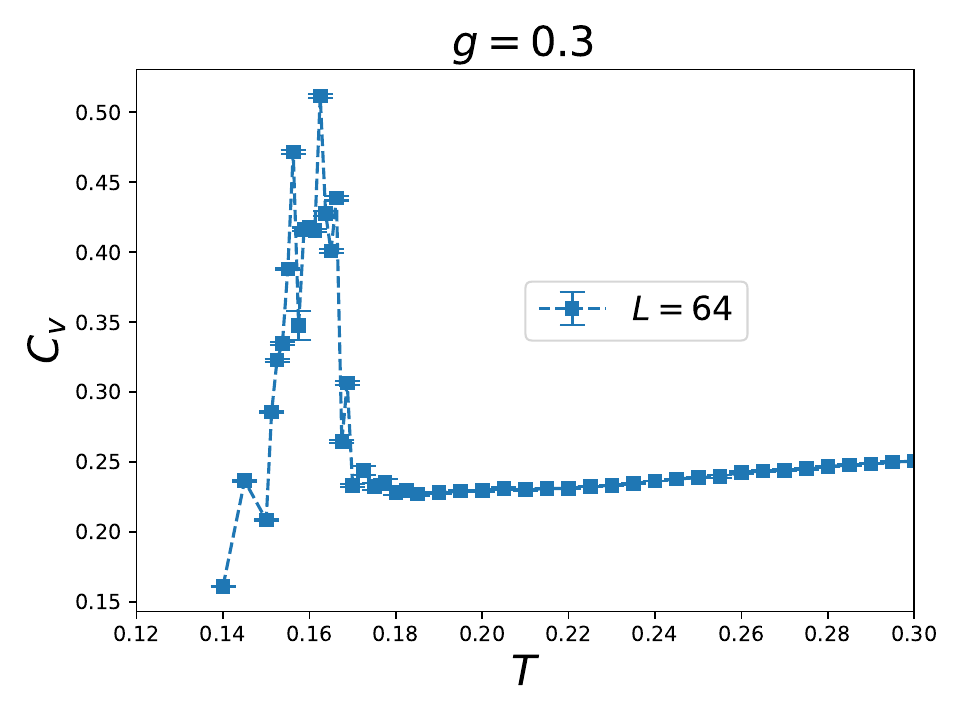}
			\includegraphics[width=0.33\textwidth]{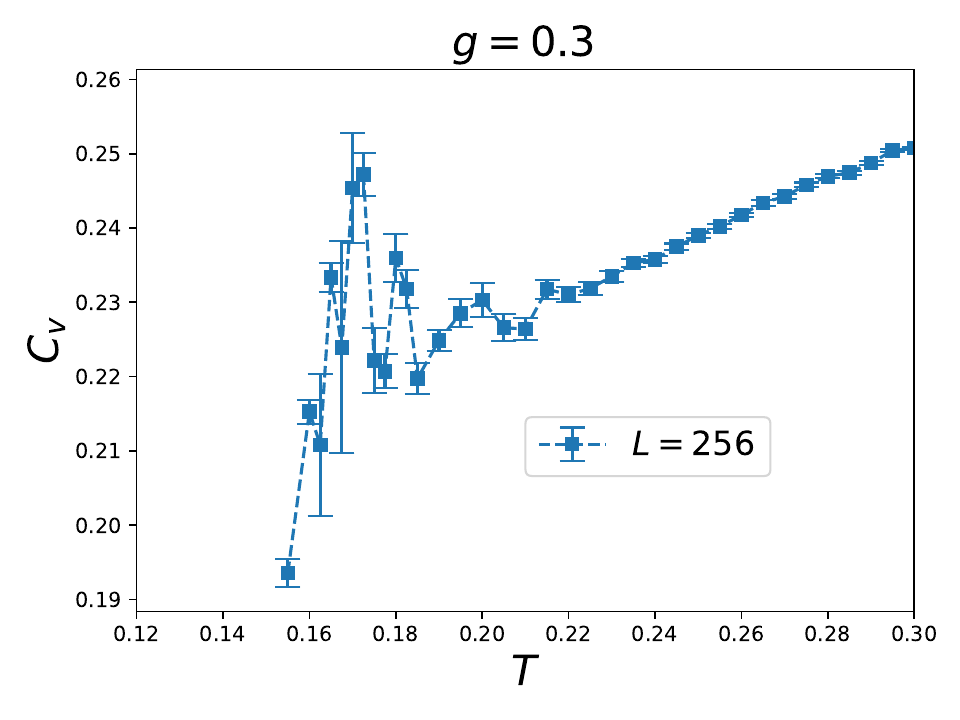}
	}
	\end{center}
	\caption{$C_v$ as functions of $T \le 0.3$ for $L = 32$ (left panel), $L=64$ (middle panel), and $L=256$ (right panel) for $g=0.3$.}
	\label{cve2}
\end{figure}

$C_v$ as functions of $T \le 0.3$ for $L =32, 64$, and 256 are depicted as the left, the middle and the right panels Fig.~\ref{cve2}, respectively. The results of fig.~\ref{cve2} indicates that for $L=32$, multiple sharp peaks do appear. Similarly, sharp peaks do show up for $L=64$ and $L=256$. However, the magnitude of these mentioned peaks diminish with increasing $L$. 

These numerical outcomes imply there is no phase transition for $g=0.3$. If it turns out finally that a phase transition occurs for $g=0.3$, then the $T_c$ must have a value much smaller than 0.14. 
The extremely large $\tau_{\text{int}}$ at low-temperature region prohibits us from getting data of $T \le 0.14$ for large $L$. 
 
We would like to point out that it is a surprise that the authors of
Ref.~\cite{Zuk19} can obtain data of $C_v$ and $E$ for $T \le 0.13$.
In particular, the data shown in Ref.~\cite{Zuk19} come with errors unnoticeable. This is not what we have found from our data. Due to this, we have conducted an investigation at $g=0.3$ with $T=0.135$ and $L=72$.

We firstly notice that all the data points of $L=72$ in Ref.~\cite{Zuk19} are
determined from one single MC simulation with total MC sweeps being $1\times 10^7$.
In particular, each calculated data is based thermalization of $2 \times 10^4$ MC sweeps and measurements of $8 \times 10^4$ MC sweeps. 

According the information outlined in the previous paragraph, for $g=0.3$ with $T=0.135$ and $L=72$, we have
carried out MC simulations with $8 \times 10^7$ MC sweeps for the thermalization. In addition, the simulations are started with
randomness-like, ferromagnet-like, and stagger-like configurations.
We would like to emphasize the fact again that even with our set up of binning the MC results before performing the measurements, we are not capable of obtaining outcomes with $\tau_{\text{int}} < 100$ at the temperature $T=0.135$ for 
$L \ge 64$.

\begin{figure}
	\begin{center}
		\hbox{
		\includegraphics[width=0.31\textwidth]{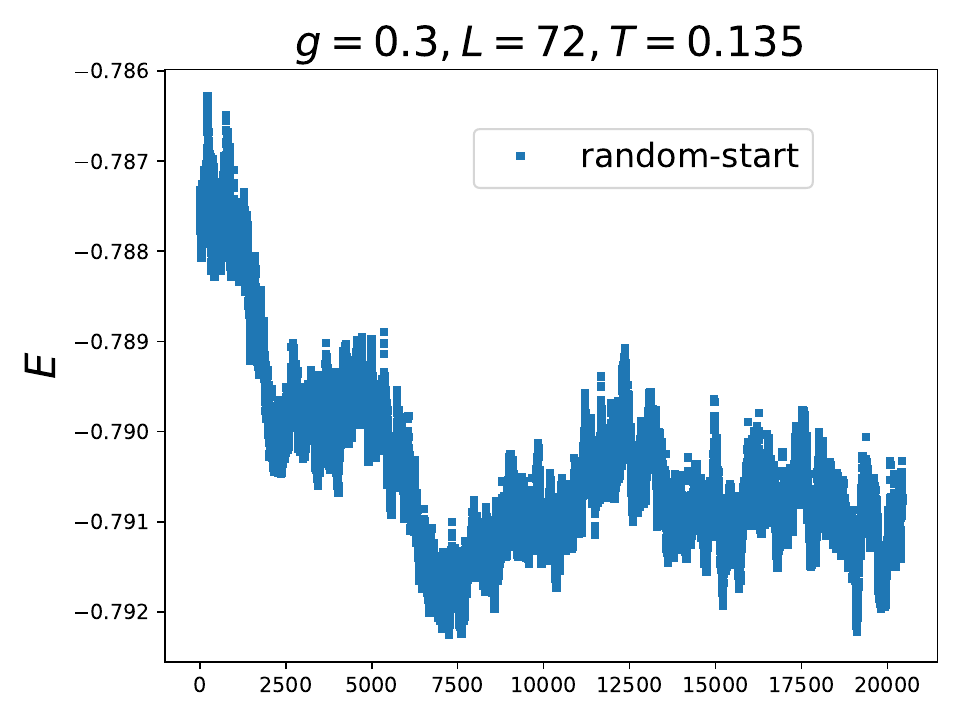}
\includegraphics[width=0.31\textwidth]{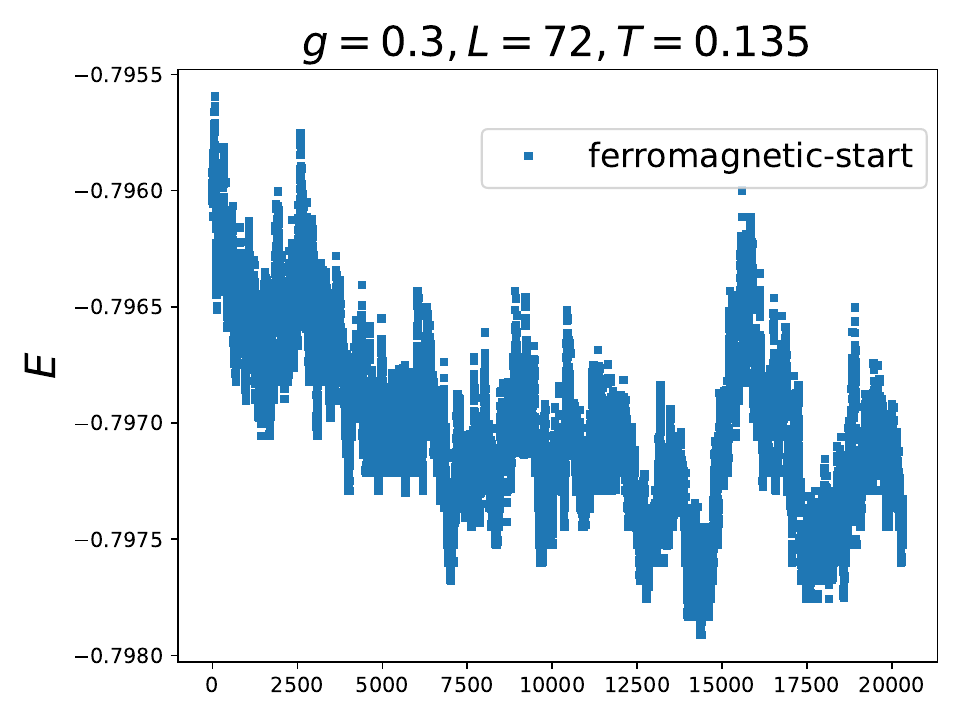}	
		\includegraphics[width=0.31\textwidth]{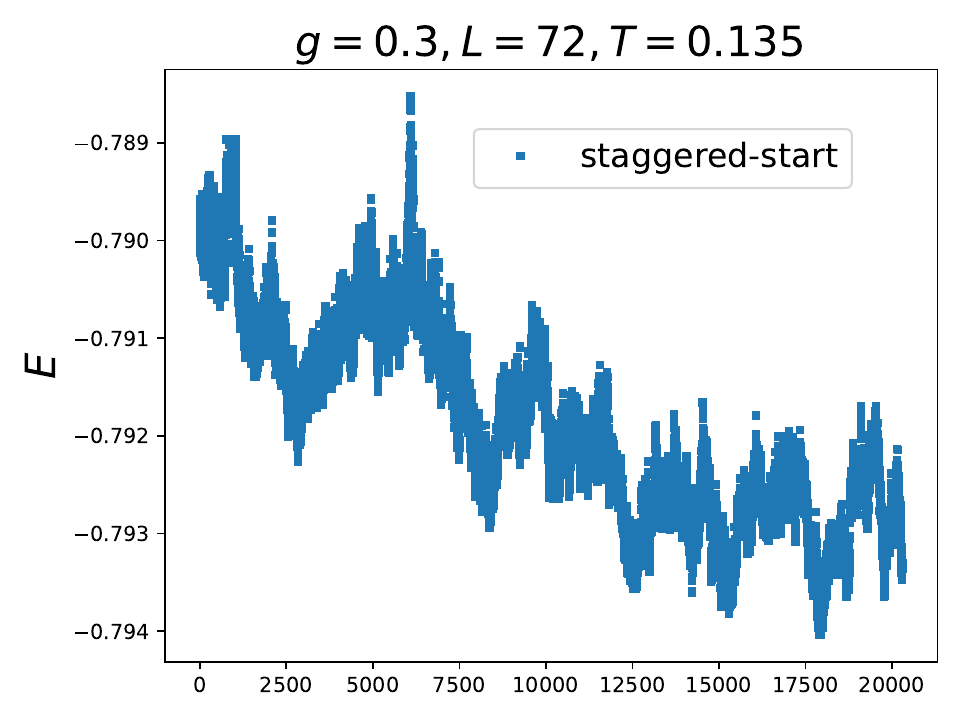}
}
		
	\end{center}
	\caption{The MC series of $E$ for $g=0.3$, $L=72$, and $T=0.135$. the left, the middle, and the right panels use randomness-like,
		ferromagnet-like, and stagger-like configurations as the start configurations for
		the Monte Carlo simulations, respectively.
	}
	\label{E13}
\end{figure}

The running histories of $E$ obtained from these mentioned investigations are represented in fig.~\ref{E13}. Each data point in every panel of fig.~\ref{E13} is determined by averaging over the outcomes of $5\times 10^4$ MC sweeps. 
In addition, the left, the middle, and the right panels use randomness-like,
ferromagnet-like, and stagger-like configurations as the start configurations for
the Monte Carlo simulations, respectively.

The outcomes in fig.~\ref{E13} indicate that either the simulations have not reached equilibrium steady states or the correlations among consecutive MC data
are extremely large. Since the total MC sweeps for obtaining each of the results of
fig.~\ref{E13} is more than $ 10^9$, one concludes that a short MC run (such as $8 \times 10^4$ as that considered in Ref.~\cite{Zuk19}) cannot truly capture the actual correlation among data. Particularly, one would expect much larger 
integrated autocorrelation lengths at temperatures $T \ll 0.135$. These outcomes
of fig.~\ref{E13} also imply the inefficiency of single spin flip algorithm when it is
applied to study the low-temperature region of the considered model.

Nevertheless, our results in conjunction with that of Ref.~\cite{Zuk19}
suggest that for $=0.3$, the scenario of no phase transition occurring is favored.

\subsubsection{ $0.24 < g < 0.3$}

From our calculations, we find that due to the abnormally large $\tau_{\text{int}}$ and (or) the extremely long equilibrium times for the $g$ lying between 0.24 and 0.3,
we are incapable of exploring the critical phenomena of these values
of $g$. Our conclusion is consistent with those explicitly (or implicitly) mentioned in Refs.~\cite{Zuk19} and \cite{Zuk21}. 
Appendix A provides related data to support this claim.

It is surprising that the critical points for the values of $g$ between 0.24 and 0.3 are presented in Ref.~\cite{Ace21}. Those $T_c$ are determined from the peaks of the related $C_v$. Since no original MC data of $C_v$ are available in Ref.~\cite{Ace21}, we are not able to
understand how the authors of Ref.~\cite{Ace21} estimate the values of $T_c$ for $0.24 < g < 0.3$. The $T_c$ for $g$ between 0.24 and 0.3 are estimated and presented in Ref.~\cite{Ace21} may be due to the fact that small system sizes ($L \sim 30$) are considered in Ref.~\cite{Ace21}.

We notice that the value of $T_c$ for $g=0.3$ shown in Ref.~\cite{Ace21} is between 0.3 and 0.5.
Since we do find a round peak near $T=0.36$, it is possible that the authors of Ref.~\cite{Ace21}
mistakenly identify this round peak as the $T_c$.

\section{Discussions and Conclusions}

In this study, we investigate the phase diagram of the 2D frustrated $J_1$-$J_2$ Ising model on the honeycomb lattice using first principles non-perturbative MC calculations. By carefully analyzing the quantities
$m$, $\chi$, $C_v$, and $E$, we conclude that the phase transitions
for $g=0.20, 0.22, 0.23$ and 0.24 are second order and belong to the 2D Ising universality class. In addition, there are phase transitions
as one moves from low-temperature region to high-temperature region
for $g=1.0, 0.8$ and 0.5. Finally, it is of high possibility that 
no phase transition occurs for $g=0.3$.

We find that $\tau_{\text{int}}$ for the considered observables are 
abnormal large at low-temperature region. This prevents us from exploring the possible critical phenomena of $0.24 < g < 0.3$.
Other approaches, such as the Tensor Network (TN), may be required
to uncover whether there are phase transitions and, if they exist, their universality class for the mentioned values of $g$.

{\it Note 1 added:} A preprint which basically considers the same model with $g \le 0.23$
has appeared in arXiv (arXiv:2509.03414). 
 	
{\it Note 2 added:} Very recently, a publication studying the full phase diagram of the considered model appears (Ref.~\cite{Azh25}).   	

\begin{figure}
	\begin{center}
		\hbox{
			\includegraphics[width=0.31\textwidth]{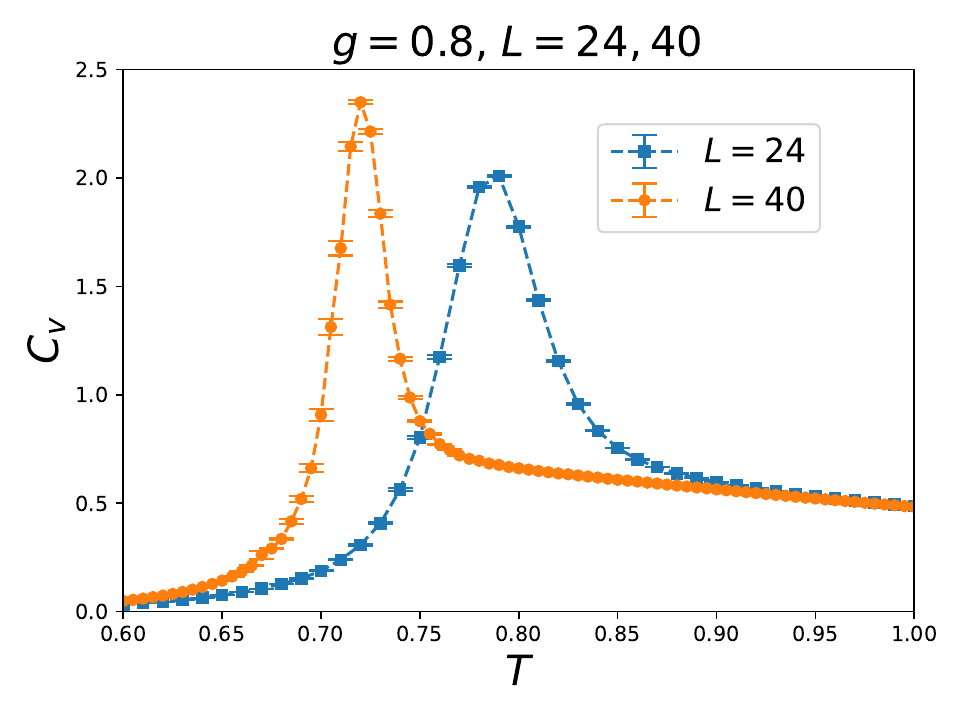}
			\includegraphics[width=0.31\textwidth]{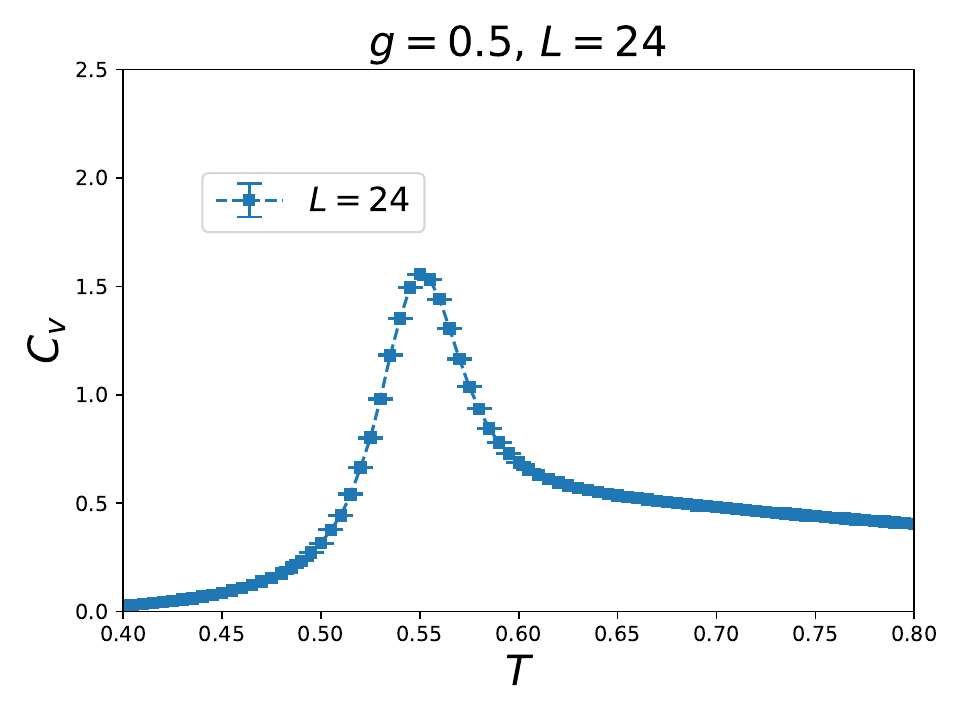}	
			\includegraphics[width=0.31\textwidth]{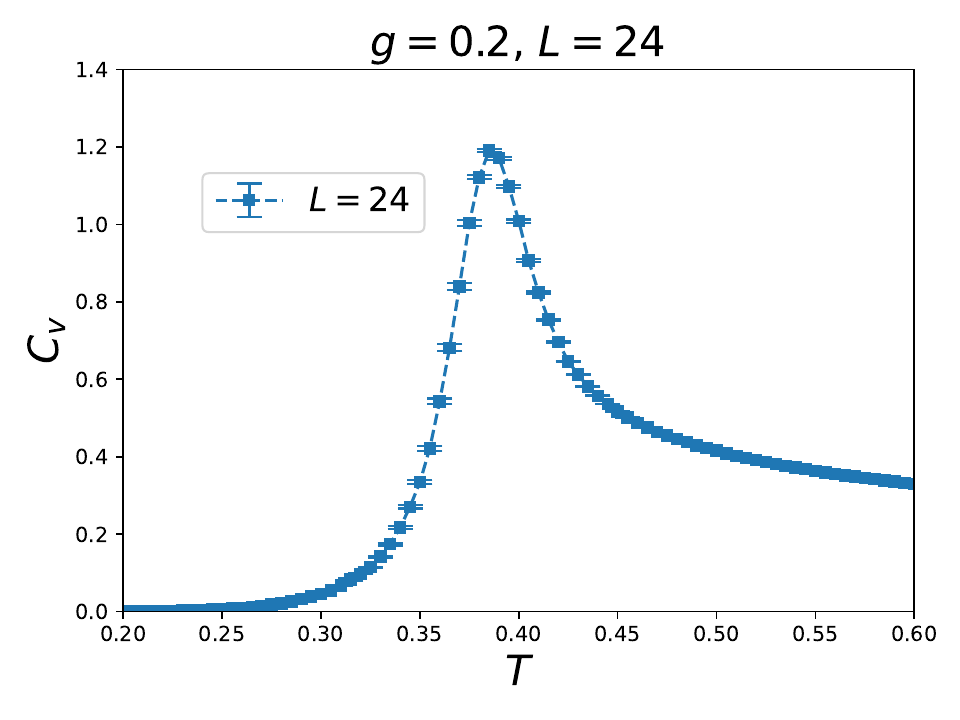}
		}
		
	\end{center}
	\caption{$C_v$ as functions of $T$ for $g=0.8$ (left),
		0.5 (middle), and 0.2. The system sizes are $L=24$ or $L=40$.
	}
	\label{comparison}
\end{figure}

The left, the middle, and the right panels of fig.~\ref{comparison} show the $C_v$
as functions of $T$ for $g = 0.8$, $0.5$, and 0.25, respectively. The system sizes are $L=24$ or $L=40$. The outcomes of $g=0.8$ and $g=0.5$ ($g=0.2$) are obtained using the stagger-like start (ferromagnet-like start). The results demonstrate in the figure agree excellently
with those given in Ref.~\cite{Azh25}. Most of the data given in Ref.~\cite{Azh25} are restricted to $L \le 24$, 40 or 100 for various $g$. Some conclusions from Ref.~\cite{Azh25} seem to disagree with ours.
The system sizes used here are much larger than those considered in Ref.~\cite{Azh25}. In addition, extremely large Monte Carlo data are used
for obtaining the results shown here. In particular, a careful analysis
by taking into account the integrated autocorrelation times and the equilibrium times are conducted for the outcomes presented in this study. 

\section*{Funding}\vskip-0.3cm
Partial support from National Science and Technology Council (NSTC) of
Taiwan is acknowledged (Grant numbers: NSTC 113-2112-M-003-014- and
NSTC 114-2112-M-003-004-).

\begin{figure}
	\begin{center}
		\vbox{
			\hbox{
				\includegraphics[width=0.45\textwidth]{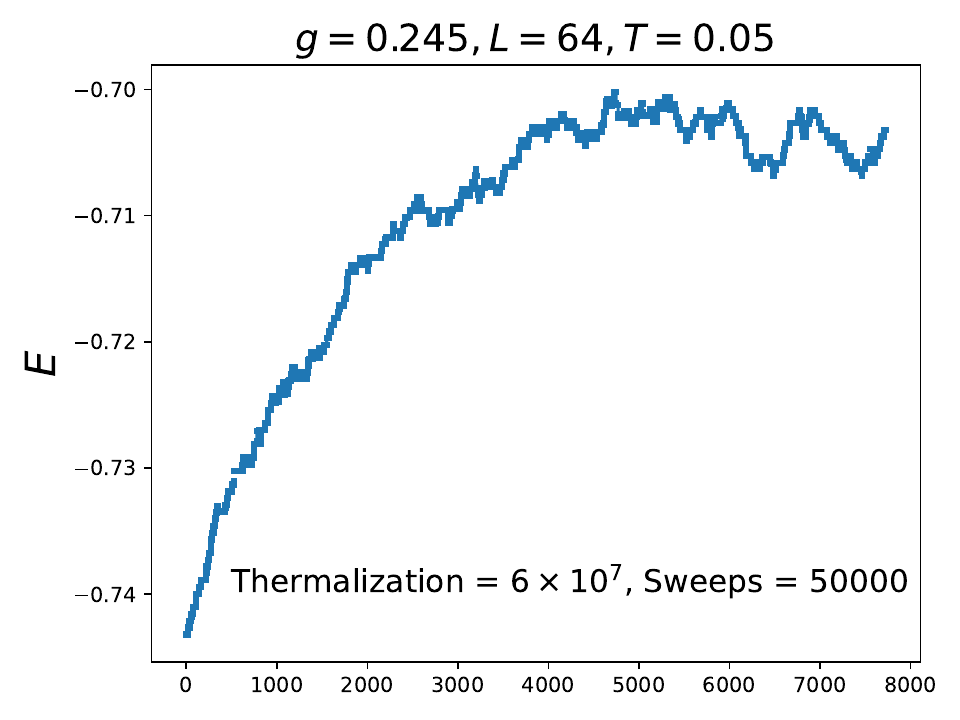}
				\includegraphics[width=0.45\textwidth]{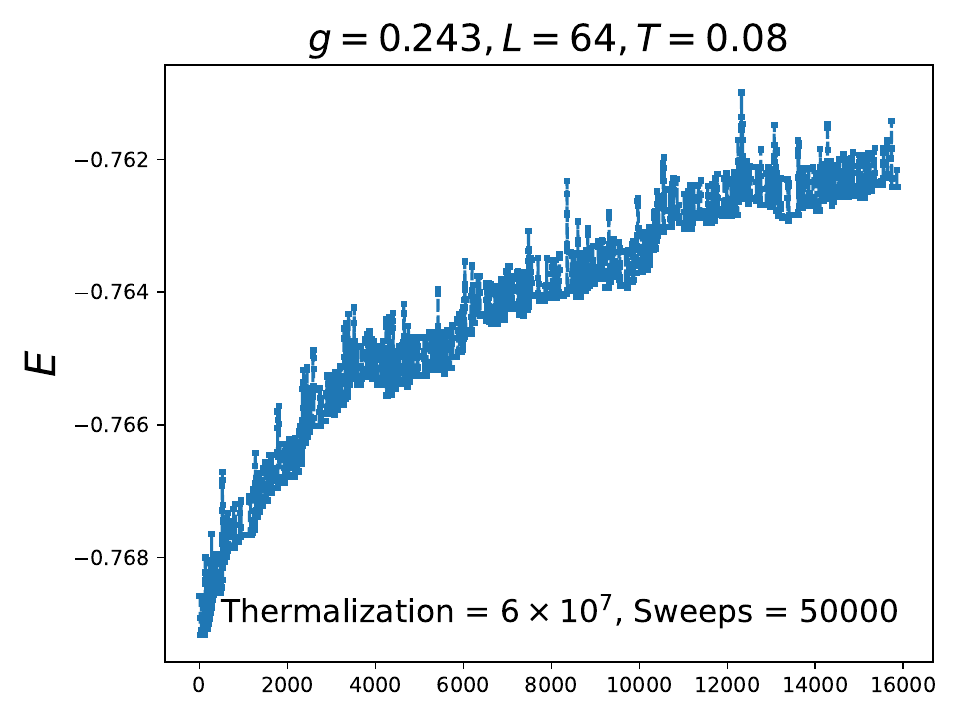}	
			}
			\hbox{
				\includegraphics[width=0.45\textwidth]{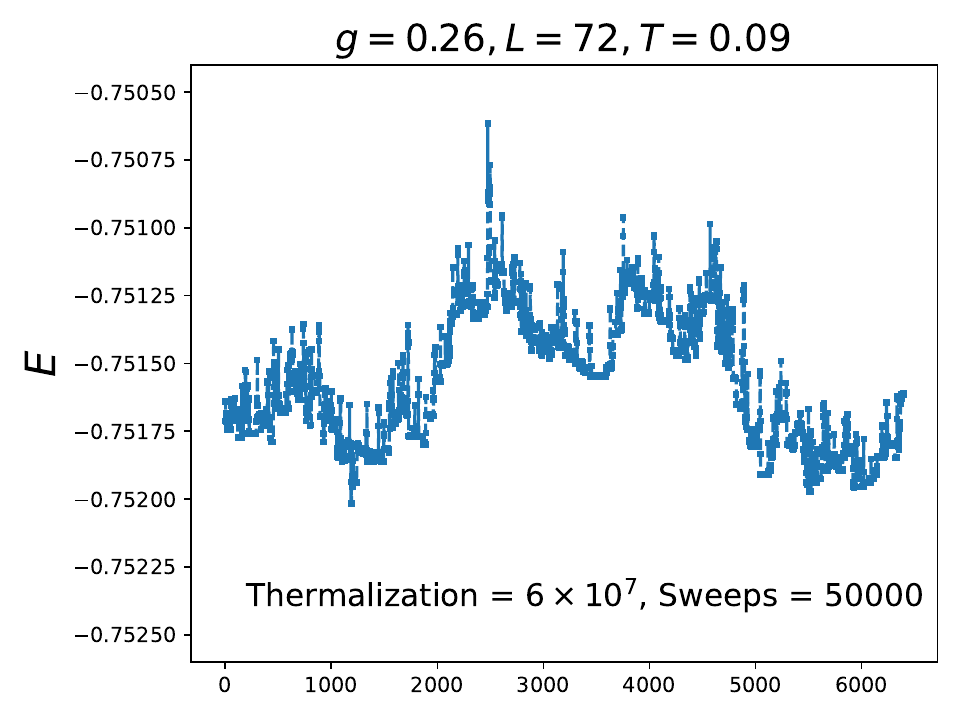}
				\includegraphics[width=0.45\textwidth]{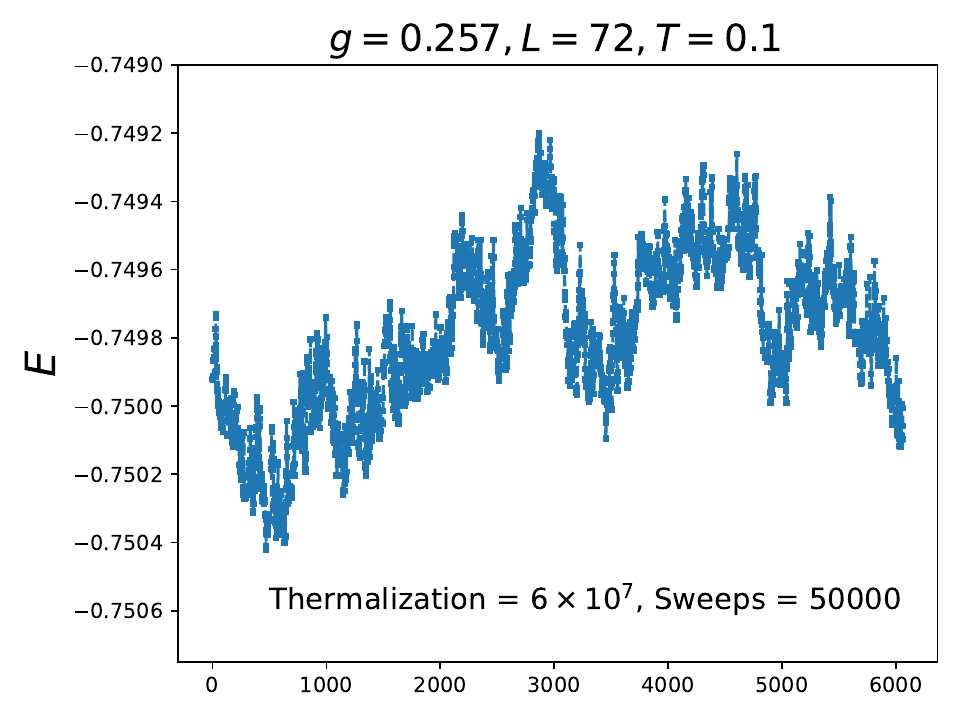}	
			}	
		}
	\end{center}
	\caption{The running histories of $E$ for several $g \in \left(0.24,0.3 \right)$ and $L=72$ ($L=64$). Each data is obtained
		by performing $6 \times 10^7$ MC sweeps for thermalization and
		averaging the outcomes of conducting 50000 MC sweeps.
	}
	\label{appendixA}
\end{figure}

\section*{Appendix A}

After performing $6 \times 10^7$ MC sweeps for the thermalization(s),
the running histories of $E$ for several values of $g$ lying between 0.24 and 0.3 with $L=72$ or $L=64$ (These are not large system sizes)
are shown in fig.~\ref{appendixA}. The value of each data point is the mean of the results of conducting 50000 MC sweeps. As can be seen in the figure, 
for those considered $g$, the corresponding equilibrium times and (or)
the integrated autocorrelation times are gigantically large.

\section*{Appendix B}
 
In our investigation, three-types of initial configurations are used to start the MC calculations. For $g = 1.0, 0.8, 0.5,$ and 0.3,
we consider stagger-start to begin the MC simulations.

\begin{figure}
	\begin{center}
		\hbox{
			\includegraphics[width=0.31\textwidth]{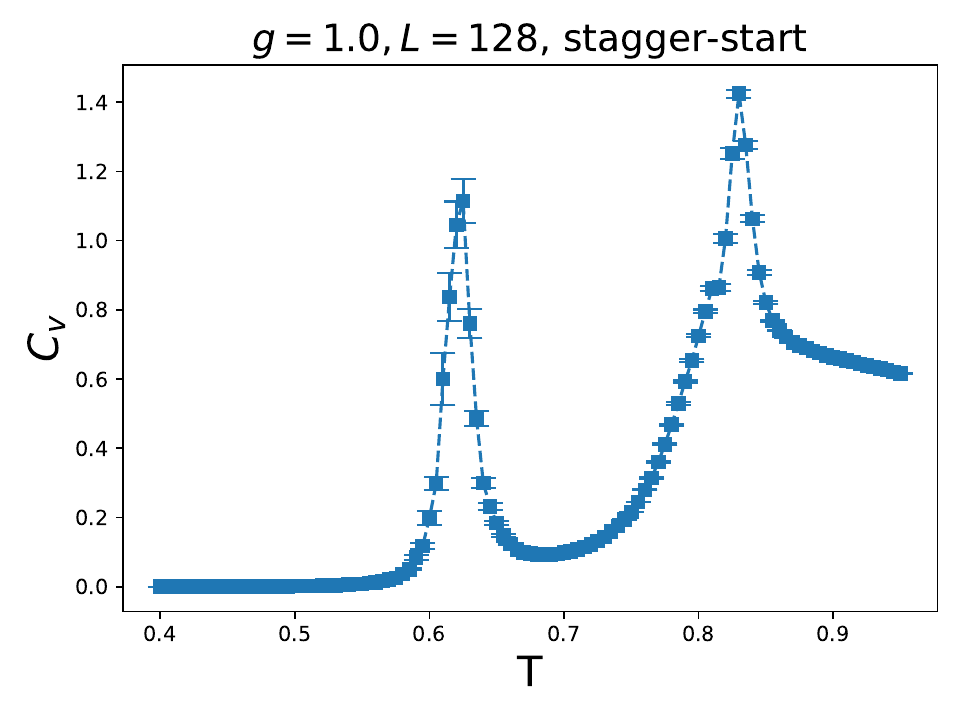}
			\includegraphics[width=0.31\textwidth]{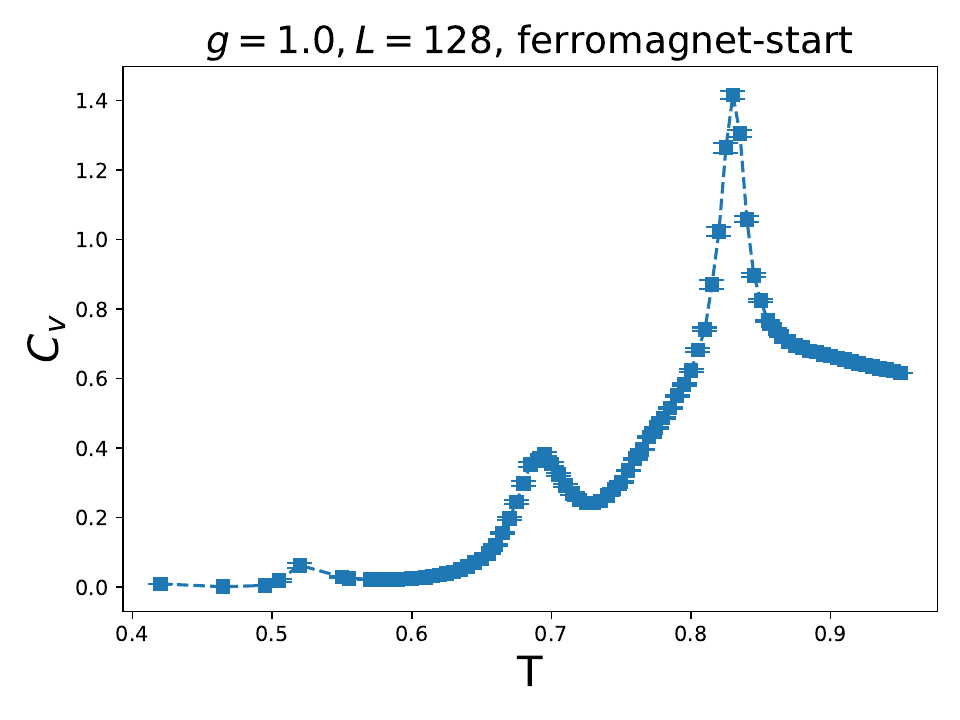}	
			\includegraphics[width=0.31\textwidth]{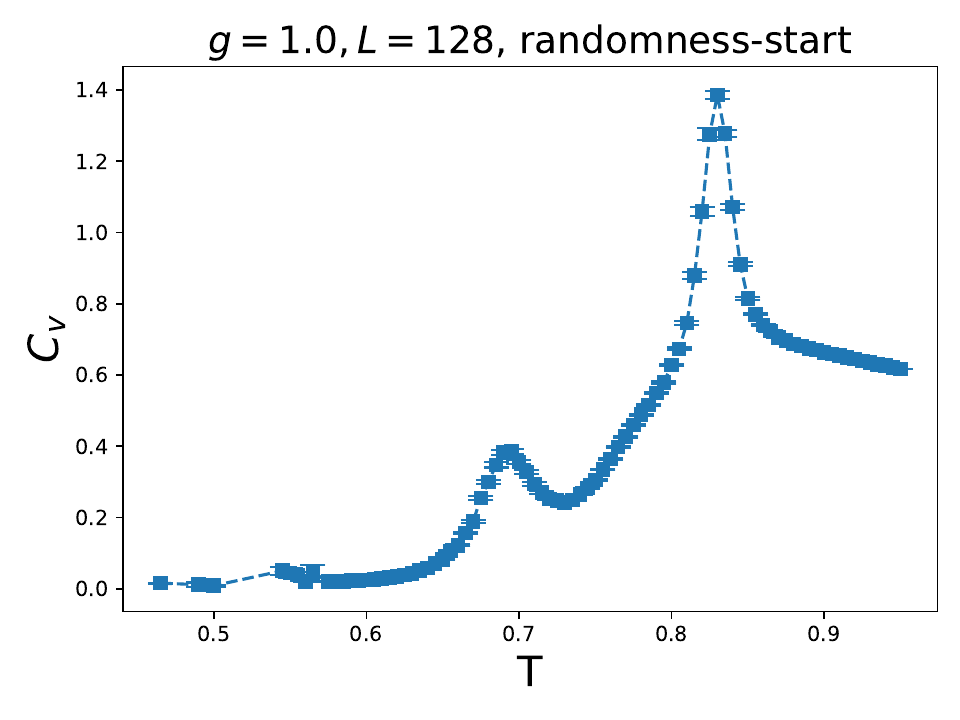}
		}
		
	\end{center}
	\caption{$C_v$ as functions of $T$ for stagger-start (left),
		ferromagnet-start (middle), and randomness-start.
	}
	\label{3typecv}
\end{figure}

\begin{figure}
	\begin{center}
		\hbox{
			\includegraphics[width=0.31\textwidth]{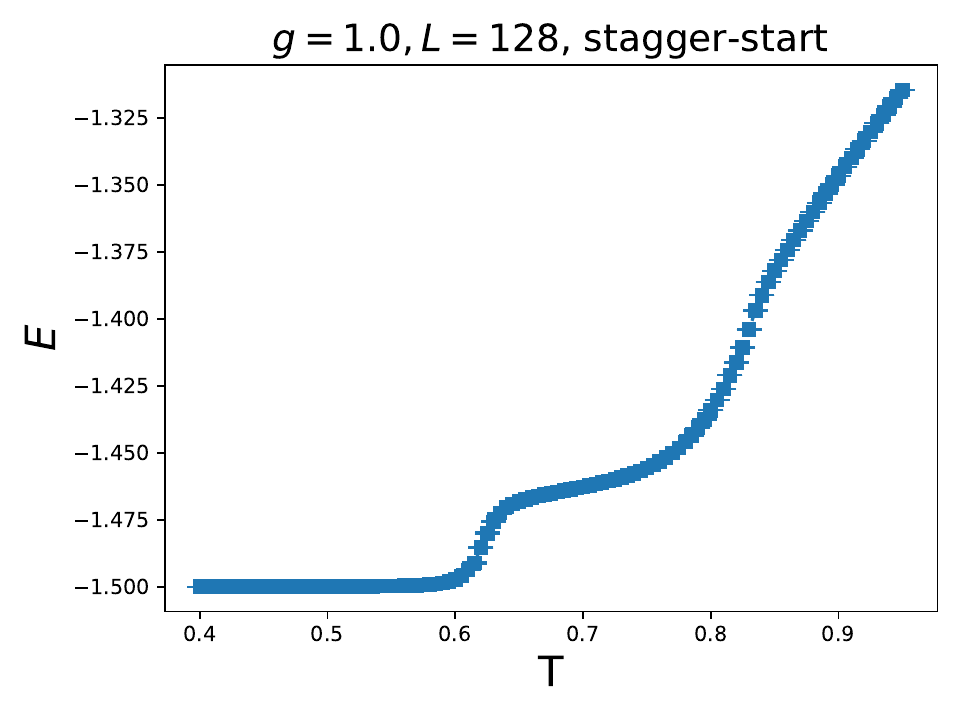}
			\includegraphics[width=0.31\textwidth]{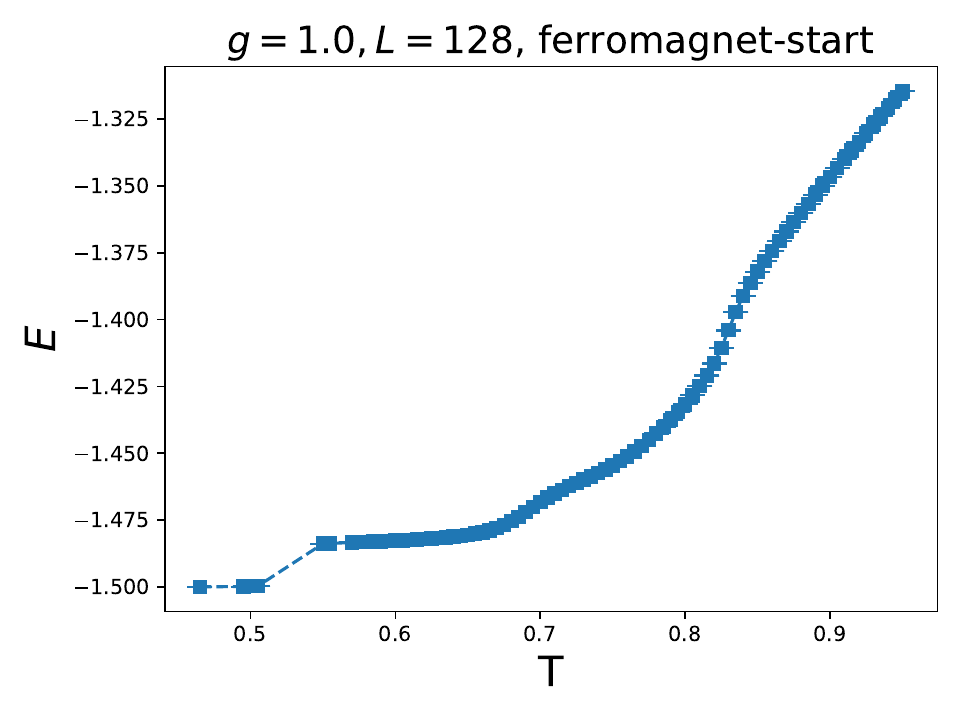}	
			\includegraphics[width=0.31\textwidth]{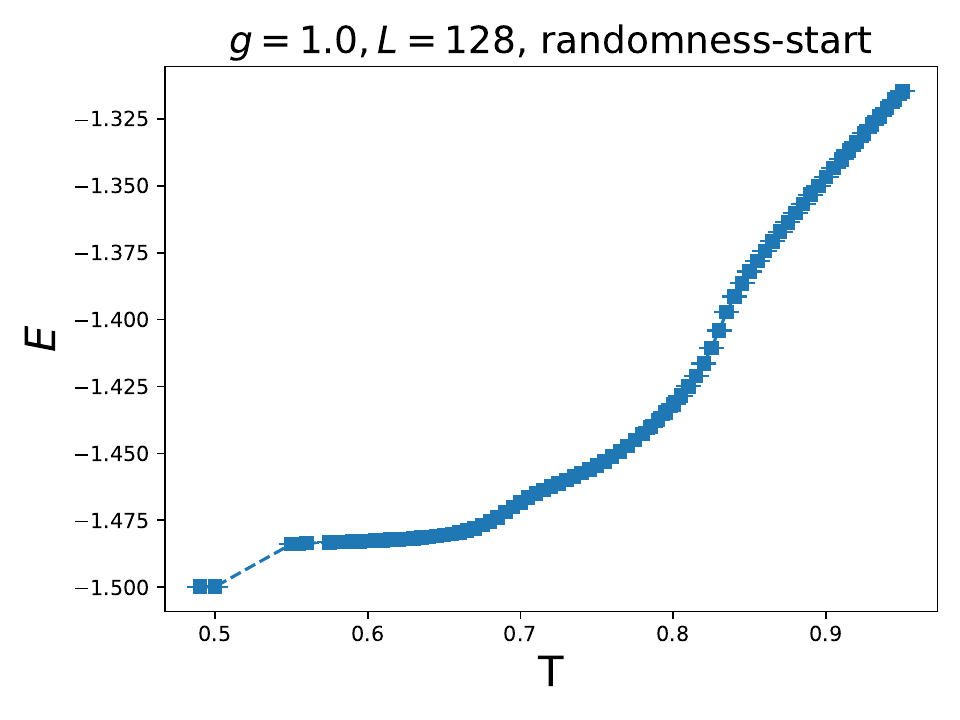}
		}
		
	\end{center}
	\caption{$E$ as functions of $T$ for stagger-start (left),
		ferromagnet-start (middle), and randomness-start.
	}
	\label{3typee}
\end{figure}

With $g = 1.0$ and $L=128$, the $C_v$ as functions of $T$ for the
stagger-, the ferromagnet-, and the randomness-start are shown
as the left, the middle, the right panels of fig.~\ref{3typecv}. 
In addition, $E$ as functions of $T$ for these three initial starts are
shown as the left, the middle, and the right panels of fig.~\ref{3typee}. 
The lowest values of $T$ considered in these simulations are $T=0.400$  
 
As can be seen from figs.~\ref{3typecv} and \ref{3typee}, in the low-$T$
region, the results associated with stagger-start are more stable and have (much) smaller integrated autocorrelation times. Since when
one approaches $g = 0.25$ from above, the magnitude of $T_c$ should become
smaller and smaller (i.e. the transitions take place at low temperatures if they exist), the use of stagger-start to begin the MC simulations
may help in obtaining more accurate estimations of $T_c$.

Surprisingly, although the $C_v$ of all the three panels of fig.~\ref{3typecv} have two peaks, the height of the peak related to the one at low temperature for the stagger-start are higher than those of the ferromagnet- and the randomness-start. The values of $T$ corresponding to these mentioned peaks between the stagger-start and the other two are different as well. Because a huge amount of data are obtained, independent written codes by the authors lead to results agreeing with each other quantitatively, careful analysis for integrated autocorrelation times are done, and the validity of the use of stagger-start is shown in the left and the right panels of fig.~\ref{comparison}, the reason behind this observation is not transparent. 

\begin{figure}
	\begin{center}
		\hbox{
			\includegraphics[width=0.31\textwidth]{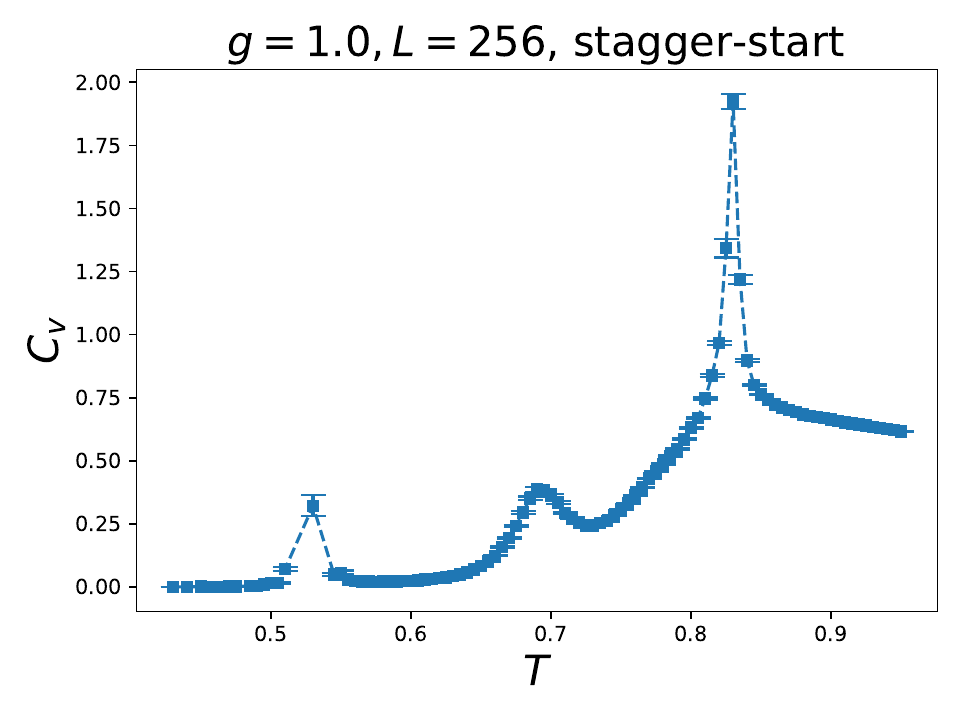}
			\includegraphics[width=0.31\textwidth]{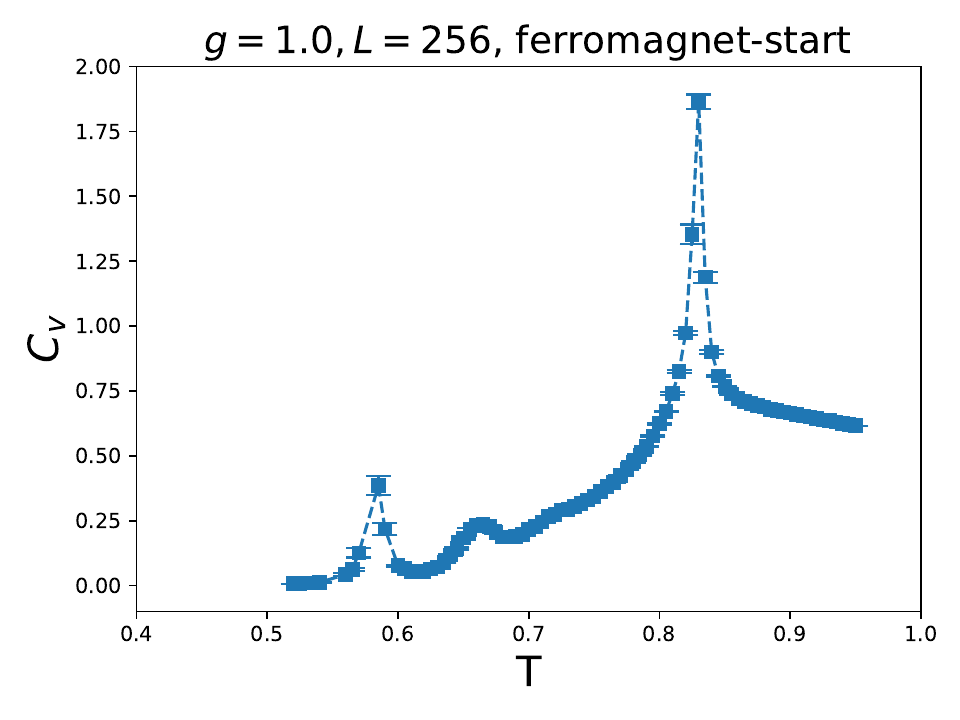}	
			\includegraphics[width=0.31\textwidth]{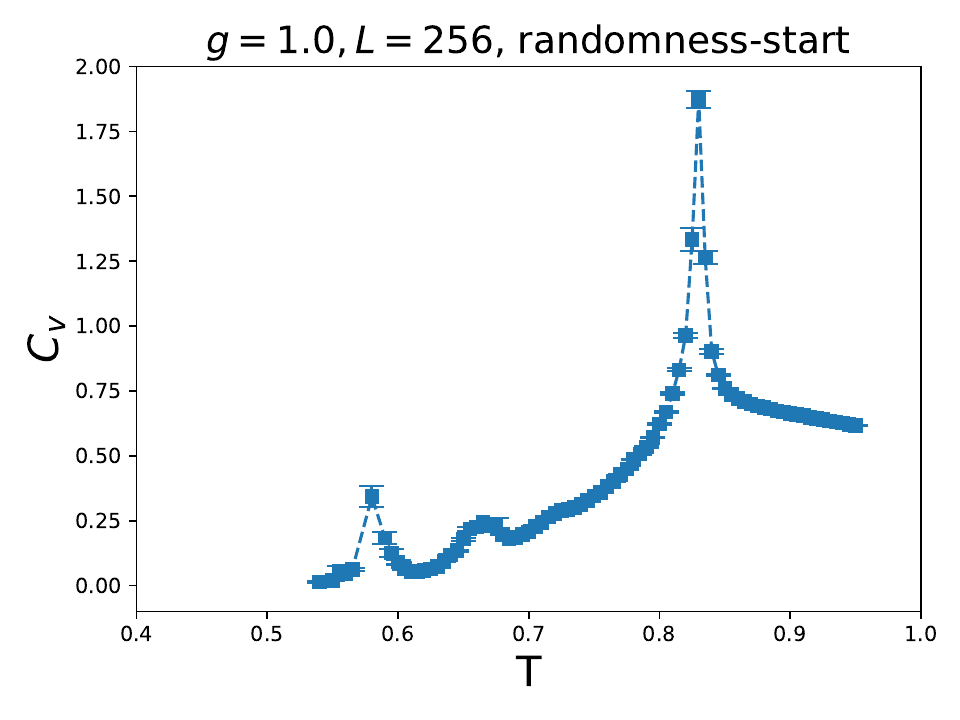}
		}
		
	\end{center}
	\caption{$C_v$ as functions of $T$ for stagger-start (left),
		ferromagnet-start (middle), and randomness-start.
	}
	\label{3typecvL256}
\end{figure}

The left, the middle, and the right panels of Fig.~\ref{3typecvL256} are $C_v$
as functions of $T$ for the stagger-, the ferromagnet-, and the randomness-start with
$L=256$. The lowest values of $T$ considered for these simulations are $T=0.400$. Clearly, in all three panels the heights of the peaks, except the ones associated with the highest temperatures, become
smaller than those related to $L=128$. In addition, in all the panels of figs.~\ref{3typecv} and \ref{3typecvL256}, the temperatures related to the most right peaks are located at about 0.85. Finally,
for each $L$, the peaks (corresponding to $T \sim 0.85$) are about the same heights as well. These outcomes imply that as $L$ increases, only the true bulk peak in $C_v$ survives. It is also noticed that for $L=256$, the integrated autocorrelation times
related to the stagger-start are smaller than the other two starts at low-$T$ region.

Due to the good features of being more stable and having smaller integrated autocorrelation times, the stagger-start is used as the initial configuration to begin the MC simulations for $g > 0.25$. In our investigation, apart from being caution about the integrated autocorrelation times, we do also pay special attention to make sure that the peaks corresponding to the fake $T_c$ are disappearing by simulating sufficient large lattices.

\begin{figure}
	\begin{center}
		\vbox{
			\hbox{
				\includegraphics[width=0.425\textwidth]{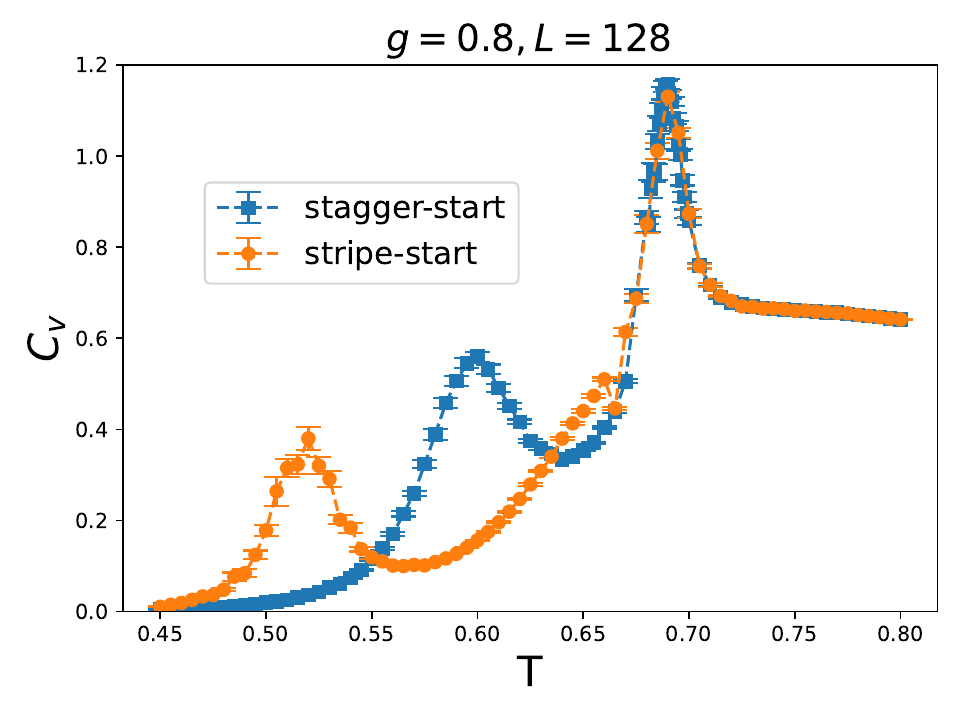}
				\includegraphics[width=0.425\textwidth]{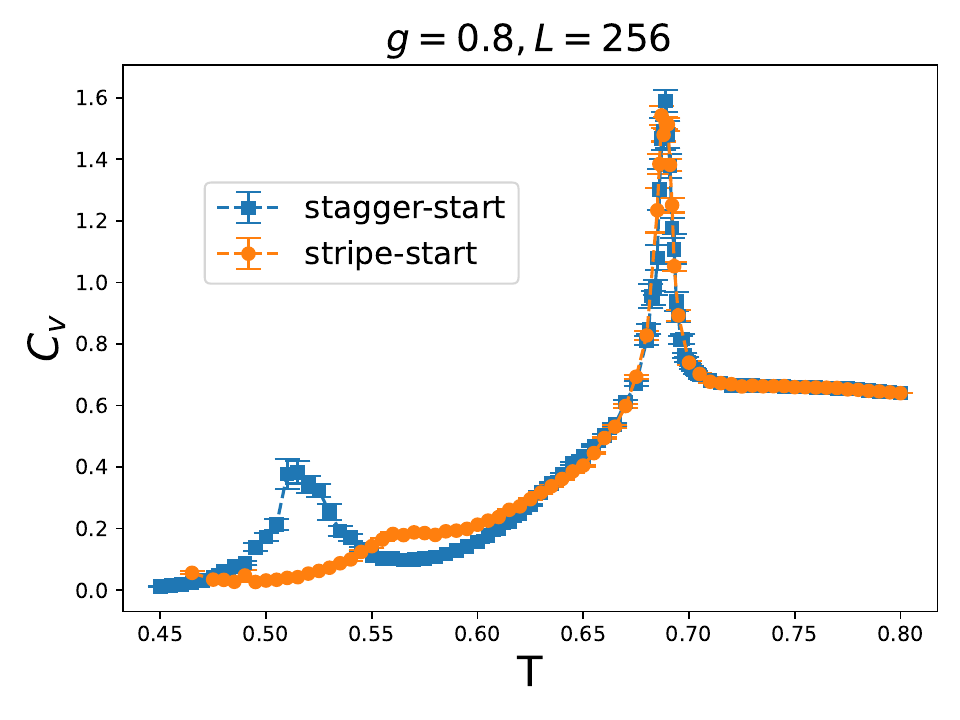}	
			}
			\hbox{
				\includegraphics[width=0.425\textwidth]{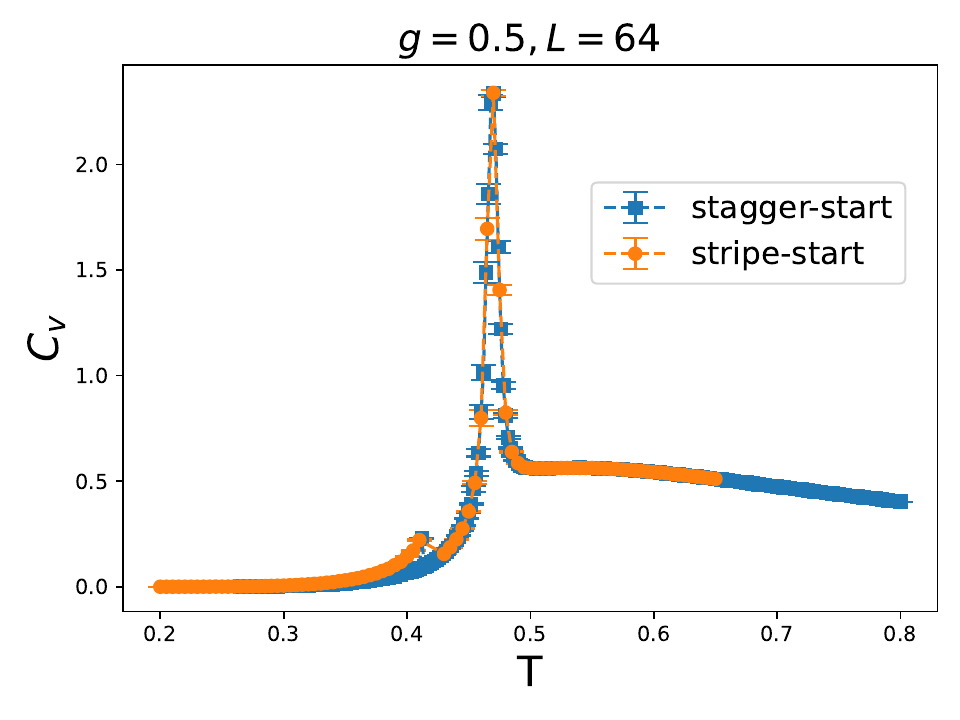}
				\includegraphics[width=0.425\textwidth]{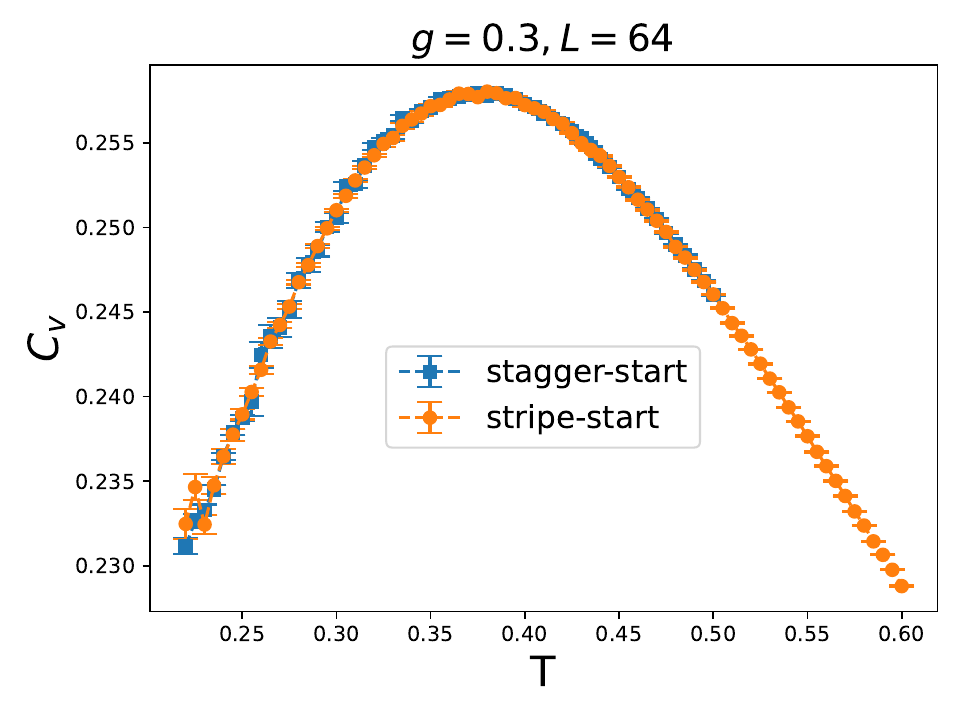}	
			}	
		}
	\end{center}
	\caption{$C_v$ as functions of $T$ for several $g$ and $L$ for both the stagger- and the stripe-start.
	}
	\label{appendixB}
\end{figure}

Finally, it should be pointed out that apart from the three configurations introduced about to begin the MC simulations, another configuration, namely the one(s) with their spins being arranged in a stripe pattern, has short integrated autocorrelation and equilibrium times at the low-$T$ region as well.
Similar to the scenario found above, for each considered $L$ the peak of $C_v$ located at the highest temperature and its height match well with those related to the stagger-start. In particular, like all the other three initial configurations, for stripe configuration, only the peak (in $C_v$)
corresponding to the highest temperature survives in the $L \rightarrow \infty$ limit if they do not vanish in this limit. Finally, similar to what's been shown in previous section(s), for $g=0.3$ and stripe-start, only round peak and no sharp peak is observed when $T \ge 0.22$, see figs.~\ref{appendixB} for the scenarios described in this paragraph.

\end{document}